\documentclass[twocolumn,aps,prl,preprintnumbers,showpacs,nofootinbib]{revtex4}
\usepackage[utf8]{inputenc}
\usepackage[english]{babel}
\usepackage{amsmath}
\usepackage{amsfonts}
\usepackage{amsmath}
\usepackage{amssymb}
\usepackage{epsfig}
\usepackage{booktabs}
\usepackage{graphics,psfrag,rotating}
\usepackage{graphicx}
\usepackage{dcolumn}
\usepackage{bm}
\usepackage{epstopdf}
\usepackage{color}
\usepackage[usenames,dvipsnames,svgnames]{xcolor}
\usepackage[colorlinks=true,
            linkcolor=red,
            urlcolor=gray,
            citecolor=blue]{hyperref}
  \usepackage{hyperref}

\newcommand{\lsim}   {\mathrel{\mathop{\kern 0pt \rlap
{\raise.2ex\hbox{$<$}}}
 \lower.9ex\hbox{\kern-.190em $\sim$}}}
\newcommand{\gsim}   {\mathrel{\mathop{\kern 0pt \rlap
{\raise.2ex\hbox{$>$}}}
\lower.9ex\hbox{\kern-.190em $\sim$}}}

\def\3nab{\tilde{\nabla}}

\def\hsp5{\hspace{5mm}}

\def\case#1/#2{\textstyle\frac{#1}{#2}}

\def\ber {\begin{eqnarray}}
\def\eer {\end{eqnarray}}
\def\bea {\begin{eqnarray}}
\def\eea {\end{eqnarray}}

\def\bc {\begin{center}}
\def\ec {\end{center}}
\def\case#1/#2{\frac{#1}{#2}}

\newcommand{\bw}{\begin{widetext}}
\newcommand{\ew}{\end{widetext}}

\newcommand{\be}{\begin{equation}}
\newcommand{\bse}{\begin{subequation}}
\newcommand{\ese}{\end{subequation}}
\newcommand{\ee}{\end{equation}}
\newcommand{\eei}{\end{eqnarray}\indent\indent}
\newcommand{\ba}{\begin{array}}
\newcommand{\ea}{\end{array}}
\newcommand{\bal}{\begin{eqnarray}}
\newcommand{\eal}{\end{eqnarray}}

\def\case#1/#2{\textstyle\frac{#1}{#2} }

\newcommand{\bver}{\begin{verbatim}}
\newcommand{\ever}{\end{verbatim}}

\begin{document}


\title{ Using the Kaniadakis horizon entropy in the presence of neutrinos to alleviate the Hubble and $ S_{8} $ Tensions  }
\author{
 Muhammad Yarahmadi\footnote{Email: Yarahmadimohammad10@gmail.com}
Amin Salehi\footnote{Email: Salehi.a@lu.ac.ir}
}
\affiliation{Department of Physics, Lorestan University, Khoramabad, Iran}

\date{\today}

\begin{abstract}
The $H_{0}$ tension stands as a prominent challenge in cosmology, serving as a primary driver for exploring alternative models of dark energy. Another tension arises from measurements of the $ S_{8} $ parameter, which is characterize the amplitude of matter fluctuations in the universe.  In this study, we address the alleviation of both the Hubble tension and $ S_{8} $ tension by incorporating Kaniadakis horizon entropy. We investigate two scenarios to explore the impact of this entropy on cosmological parameters. In the first scenario, utilizing modified Friedmann equations through Kaniadakis entropy, we  estimate the values of $H_{0}$ and $ S_{8} $. In the subsequent scenario, we introduce the neutrino term and assess its effect on mitigating the Hubble and $ S_{8} $ tensions. Our findings reveal that when considering the first scenario, the results closely align with Planck's 2018 outcomes for Hubble and $ S_{8} $ tensions. Moreover, with the inclusion of neutrinos, these tensions are alleviated to approximately 2$\sigma$, and the $ S_{8} $ value is in full agreement with the results from the KiDS and DES survey. Furthermore, we impose a constraint on the parameter $K$ in each scenario. Our analysis yields $K = 0.12\pm 0.41$ for Kaniadakis entropy without neutrinos and $K = 0.39\pm 0.4$ for the combined dataset considering Kaniadakis entropy in the presence of neutrinos. We demonstrate that the value of K may be affected by neutrino mass, which can cause energy transfer between different parts of the universe and alter the Hubble parameter value.
\end{abstract}


%
%


\maketitle
.

\section{Introduction}
By directly probing the expansion, Riess et al. (1998) \cite{1} and Perlmutter et al. (1999) \cite{2} recently observed that the Universe has entered an epoch of accelerated expansion, providing direct evidence for dark energy. These groups conducted observations of the apparent magnitudes of several type Ia supernovae. If the Universe is undergoing an accelerated rate of expansion, the energy density component responsible must exert a negative pressure.

Although gravity is a known universal force in nature, understanding its origin has long been a mystery. Einstein proposed that gravity is the curvature of space, considering it an apparent phenomenon describing the dynamics of space. In recent decades, scientists have made numerous attempts to unveil the nature of gravity. One prospective avenue explored in recent years is the study of space-time thermodynamics, revealing that Einstein's equations of general relativity are, in fact, the same equation of state of space-time. By considering the equation $\delta Q = T\delta S$ and the entropy relation, it is possible to demonstrate the equivalence between the field equations and the first law of thermodynamics.

These studies can be extended to a cosmological perspective, leading to the derivation of Friedman's equations from the first and second laws of thermodynamics. However, the origin of general relativity remains unclear from the standpoint of statistical mechanics. It is crucial to note that demonstrating the equivalence between the Friedman equation and the first law of thermodynamics $dE=T_h dS_h + dW$ at the apparent horizon requires considering the entropy of the black hole in any gravitational theory.

Furthermore, we acknowledge that the entropy associated with the black hole horizon is modified by the inclusion of quantum effects. As a result, various types of quantum corrections to the area law have been introduced, with one intriguing case being the generalized entropy known as Kaniadakis entropy \cite{3,4}.

This is a one-parameter generalization of the classical Boltzmann-Gibbs-Shannon entropy, arising from a coherent and self-consistent relativistic statistical theory. It preserves the basic features of standard statistical theory and recovers it in a particular limit. This corrective term appears in a model of entropic cosmology and can simultaneously satisfy the recent accelerated inflation and expansion of the universe.
The $H_{0}$ tension is one of the main problems in cosmology and is the single biggest motivator for the investigation of alternative models of dark energy.

The value obtained for Hubble's constant using supernova observations, similar to what Edwin Hubble did, exhibits a substantial difference from the value obtained using cosmic background radiation. Despite technological advancements and increasing measurement accuracy, contrary to expectations of a decrease, the magnitude of this difference has increased.
 There are two methods to obtain Hubble's constant. The first method, known as the direct measurement method.
 
 The direct measurement of the Hubble constant ($H_{0}$) involves measuring the apparent recession velocities of galaxies and other astronomical objects and determining their distances. One of the significant challenges in direct measurements of 
$H_{0}$ is obtaining accurate and precise distance measurements to galaxies.

The second method involves using the Cosmic Microwave Background (CMB) sound peaks with constraints for the cosmological model. While CMB observations offer significant information about cosmological parameters, the available data restrict the combination of $H_0$ with other parameters. Additional assumptions or data must be provided to derive the Hubble constant. One possible assumption is that the universe is perfectly flat (i.e., $\Omega_{K} = 0$). In this scenario, measurements of the CMB power spectrum with the WMAP and Planck satellites enable the determination of the Hubble constant.

Despite direct and indirect measurements of Hubble's constant yielding two different values, the tension has increased to 5.3$\sigma$ as measurements have become more accurate in recent years, contrary to expectations. This difference appears unrelated to measurement errors, suggesting the potential need for new physics beyond $\Lambda$CDM to accurately describe the universe.

Physicists have proposed various models to address Hubble's tension, though none have completely resolved it. We will describe some categories of these models, with reference to the article \cite{5} in this category:

1. Models where the equation of state of dark energy is either $ \omega >   -1 $ or $ \omega <   -1 $ \cite{6,7,8}.

2. Models considering a primordial dark energy component (EDE) that exists at $ z > 3000 $ and then disappears (\cite{9,10,11,12,13,14,15,16,17}).

3. Models considering the interaction between dark energy and dark matter, beyond gravitational interaction \cite{18,19,20,21,22}.

4. Models modifying the history of recombination and re-ionization \cite{23}.

5. Models modifying gravity \cite{24,25,26}.

6. Models considering decaying dark matter \cite{27,28}.

7. Models considering interacting neutrinos \cite{29,30}.

In direct measurement, to determine the Hubble constant, it is necessary to obtain the velocity of an object through spectral analysis while simultaneously accounting for its distance or luminosity. In practice, the selected object must be situated at a sufficient distance so that its motion is primarily driven by the expansion of the universe. Specifically, the expansion velocity of an object is directly proportional to its distance from Earth, while other forms of velocity, such as those caused by gravitational forces exerted by surrounding objects, are negligible in comparison.

This calibration facilitates the standardization of magnitudes for more distant supernovae in the Hubble flow. Using this method, the Supernova $ H_0 $ Equation of State (SH0ES) collaboration reports a value of $H_{0}= 74.03 \pm 1.42 \, \mathrm{kms^{-1} Mpc^{-1}}$. The release of new Pantheon data marks a significant development in cosmology, particularly in the study of Type Ia supernovae (SnIa). The catalog includes a total of 1701 Type Ia supernovae observed over a specific redshift range denoted as $0.001 < z < 2.3$. Redshift measures how much light from a distant object has been stretched as the universe expands, covering a broad span of cosmic history.

 Another tension arises from measurements of the $\sigma_{8}$ parameter. The parameter \( S_8 \) is a cosmological parameter used to quantify the amplitude of matter density fluctuations in the universe. It is defined as the root mean square amplitude of mass fluctuations within a sphere of radius \( 8h^{-1} \) Mpc, where \( h \) is the reduced Hubble constant.
 	
 	Mathematically, \( S_8 \) is expressed as: 	$ S_8 = \sigma_8 \sqrt{\frac{\Omega_m}{0.3}}$ 
 	Here, \( \sigma_8 \) is the amplitude of matter fluctuations on scales of \( 8h^{-1} \) Mpc, and \( \Omega_m \) is the density parameter for matter in the universe. Amplitude of Fluctuations: \( S_8 \) essentially tells us about the amplitude or strength of the density fluctuations in the matter distribution of the universe. Cosmological Structure Formation: The parameter is crucial in the context of cosmological structure formation. Fluctuations in the density of matter seed the formation of cosmic structures such as galaxies and galaxy clusters.
 	Comparison with Observations: Observations of large-scale structures in the universe, such as galaxy surveys or cosmic microwave background (CMB) measurements, can be compared with theoretical predictions based on cosmological models. \( S_8 \) provides a convenient way to parameterize and compare these predictions. Cosmological Constraints: The value of \( S_8 \) is influenced by various cosmological parameters, including the matter density (\( \Omega_m \)) and the amplitude of initial density fluctuations (\( \sigma_8 \)). By measuring \( S_8 \) and comparing it with observations, constraints on these cosmological parameters can be derived.

The value inferred from the Planck CMB measurement is $S_{8} = 0.832 \pm 0.013$. However, there is a 2$\sigma$ tension with measurements of $\sigma_{8}$ coming from galaxy clusters and weak lensing. 
We also consider cases including Gaussian priors on $S_8$ as measured by KiDS-1000x\{2dFLenS+BOSS\} ($S_8=0.766^{+0.02}_{-0.014}$) ~\cite{51} and DES-Y3 ($S_8=0.776\pm0.017$)~\cite{50}.

\section{ Kaniadakis horizon entropy}
\label{model}
The Kaniadakis entropy serves as a valuable tool for the examination of the statistical behavior of cosmic systems, such as Cosmic Microwave Background (CMB) radiation. It proves useful in modeling CMB temperature fluctuations and exploring their implications for cosmology.

Kaniadakis statistics, also known as K-statistics, presents a generalization of Boltzmann–Gibbs statistical mechanics. It is founded on a relativistic extension of the classical Boltzmann–Gibbs–Shannon entropy, commonly referred to as Kaniadakis entropy or K-entropy.

Kaniadakis entropy retains the fundamental features of standard statistical theory and converges to it under specific limits. It represents a one-parameter generalization of the classical Boltzmann-Gibbs-Shannon entropy \cite{31,32}.

 This corrective term appears in a model of entropic cosmology and can simultaneously satisfy the recent accelerated inflation and expansion of the universe. In particular, Kaniadakis entropy is given
 by
 \begin{eqnarray}
 	S_{K}=- k_{_B} \sum_i n_i\, \ln_{_{\{{\scriptstyle
 				K}\}}}\!n_i  ,
 \end{eqnarray}
 with $k_{_B}$ the Boltzmann constant.  We introduce
 \begin{eqnarray}
 	{\ln _k} \equiv \frac{{{x^k} - {x^{ - k}}}}{{2k}},
 \end{eqnarray}

 where k is the  Kaniadakis parameter. The kaniadakis parameter, denoted as \( k \), is a dimensionless parameter used in the context of the kaniadakis statistics, which is an extension of standard statistical mechanics. This parameter plays a crucial role in characterizing the departure from conventional statistical behavior. The statistical mechanics that we are accustomed to, often associated with the Boltzmann-Gibbs entropy, is retrieved as a special case when \( k \) tends towards zero.
 
 To provide a bit more detail, the kaniadakis statistics were introduced as a generalization of the standard statistical framework to accommodate non-extensive systems. Such systems exhibit behaviors that cannot be adequately described by the classical Boltzmann-Gibbs statistics. In particular, the kaniadakis distribution is derived by maximizing the entropy under certain constraints, yielding a modified form that encompasses a broader range of physical scenarios.
 
 When \( k \) is within the range \( -1 < k < 1 \), it signifies a departure from the classical statistical behavior. As \( k \) approaches zero, the kaniadakis statistics converge towards the standard Boltzmann-Gibbs statistics, indicating the recovery of traditional statistical mechanics.Also,  $K\rightarrow 0$ recovers the standard Bekenstein-Hawking entropy, namely $S_{K\rightarrow 0}=S_{BH}$.

We trivially verify that $ S_k=S_{-k} $. Consistently, we also verify
\begin{equation}
	{\ln _k}x = \frac{{{{\ln }_q}x + {{\ln }_q}\frac{1}{x}}}{2} = \frac{{{{\ln }_q}x + {{\ln }_{2 - q}}x}}{2},
\end{equation}
 with $ q=1+k $, hence
 \begin{equation}
 	{\ln _k}x = \frac{{{{\ln }_{1 + k}}x + {{\ln }_{1 - k}}x}}{2}.
 \end{equation}
 Consequently, the definition (2) implies

  \begin{equation}
{S_k} = \frac{{{S_{1 + k}} + {S_{1 - k}}}}{2}.
 \end{equation}
Equivalently, kaniadakis entropy    can be expressed as
\cite{33,34,35,36,37,
	38,
	39,40}
\be \label{kstat}
S_{k} =-k_{_B}\sum^{W}_{i=1}\frac{P^{1+k}_{i}-P^{1-k}_{i}}{2k},
\ee
with  $P_i$ the probability of a system to be in  a specific microstate and $W$
the total configuration number.
 Hence,
 for the black hole application of kaniadakis entropy we
 obtain
 \be \label{kentropy}
 S_{k} = \frac{1}{k}\sinh{(k S_{BH})},
 \ee

 \section{Modified Friedmann equations through kaniadakis entropy}
 In this section, we first investigate the alleviation of the Tensions with the use of the Modified Friedmann equations through kaniadakis entropy\cite{41}. Then, we discrete the $ \rho_{m} $ and $ p_{m} $ to ${\rho _m} = {\rho _b} + {\rho _c} + {\rho _\nu } $ and $ {p_m} = {p_b} + {p_c} + {p_\nu } $ in field equations and best fit the kaniadakis parameter (k), after that we investigate the effect of considering neutrinos term in $ \Omega_{m} = \frac{\rho_{m}}{3H^{2}}$ where $ \Omega_{m}= \Omega_{b}+ \Omega_{\nu}+ \Omega_{c} $ to alleviate the  Hubble Tension and $ \sigma_{8} $  Tension.
 We start from the modified Friedmann equation in flat case, \cite{41}
 \begin{eqnarray}
 	\label{FR1}
 	&&H^2 = \frac{8\pi G}{3}(\rho_{m}+\rho_{DE})\\
 	&&\dot{H} = -4\pi G(\rho_{m}+p_{m}+\rho_{DE}+p_{DE}),
 	\label{FR2}
 \end{eqnarray}
 where ${\rho _m} $ is matter density  and $ {p_m} $ denotes the pressure of matter and  ${\rho _{DE}}$ and  ${p _{DE}}$ act as dark enegy density and dark energy pressure.  The dark energy sector is defined as\cite{41}
 \begin{eqnarray}
 	&&
 	\!\!\!\!\!\!\!\!\!\!\!\!\!\!\!\!\!\!\!\!\!\!\!\!\!\!\!\!\!\!\!\!\!\!
 	\rho_{DE}=\frac{3}{8\pi G}\left \{\frac{\Lambda}{3}+H^{2}\left [1-
 	\cosh{\left(k
 		\frac{\pi}{GH^2}\right)}\right ]\right.\nonumber\\
 	&& \!\!\!\!\!\!\!  \!\!\!\!\!\!\!\!\! \left. +\frac{k\pi}{G}
 	\text{shi}{\left(k \frac{\pi}{GH^2}\right)} \right \},
 	\label{rhoDE1}
 \end{eqnarray}
 \begin{eqnarray}
 	&& \!\!\!\!\!\!\!\!\!\!
 	p_{DE}=-\frac{1}{8\pi G}\Big \{\Lambda +(3H^{2}+2\dot{H})\left [1-
 	\cosh{\left(k
 		\frac{\pi}{GH^2}\right)}\right ] \ \nonumber\\
 	&&
 	\ \ \ \ \,
 	+\frac{3k\pi}{G} \text{shi}{\left(k \frac{\pi}{GH^2}\right)} \Big
 	\}.
 	\label{pDE1}
 \end{eqnarray}
 Hence, with the effective dark energy density
 and pressure at hand, we can define the equation-of-state parameter for the
 effective dark energy sector as
 \begin{eqnarray}
 	\label{wDE}
 	&&
 	\!\!\!\!\!\!\!\!\!\!\!\!\!\!\!\!\!\!\!\!\!\!
 	w_{DE}\equiv\frac{p_{DE}}{\rho_{DE}}=-1-
 	2\dot{H}\left [1-\cosh{\left(k
 		\frac{\pi}{GH^2}\right)}\right ] \nonumber\\
 	&&
 	\ \ \ \ \ \ \ \ \
 	\cdot
 	\left\{\Lambda +3H^{2}\left [1-\cosh{\left(k
 		\frac{\pi}{GH^2}\right)}\right
 	] \right.
 	\nonumber\\
 	&&
 	\ \ \ \ \ \ \ \ \ \ \ \
 	\left.
 	+\frac{3k\pi}{G} \text{shi}{\left(k
 		\frac{\pi}{GH^2}\right)}\right\}^{-1}.
 \end{eqnarray}

Where the function $\text{shi}{(x)}$ is generally defined as $\text{shi}{(x)} = \int^{x}_{0}{\frac{\sinh(x')}{x'}dx'}$ which is a mathematical odd function of $x$ with no discontinuity. According to equations (8),(9),  we introduce the following new variables,
\begin{align}
	\Omega_{m}  = \frac{{{\rho _ m}}}{{3{H^2}}},	\Omega_{DE}  = \frac{{{\rho _{DE} }}}{{3{H^2}}},
\end{align}
	were $ \Omega_{\rm m} $ is baryon density, and $ \Omega_{DE} $  denotes as dark energy density.
Now, we can derive the following autonomous equations as:
\begin{align}
	\begin{split}
		&\Omega_{m} ' =  - 3\Omega_{m}  - 2\Omega_{m} \frac{{\dot H}}{{{H^2}}},\\
		&\Omega_{DE} ^{'} =  - 3\Omega_{DE} (1 + {\omega _{DE}}) - 2\Omega_{DE} \frac{{\dot H}}{{{H^2}}},
	\end{split}	
\end{align}

where prime denotes variation with respect to $ N $ and $ N = \rm ln a$.  It is clear that in the case where $k=0$, the generalized Friedmann equations
(10),(11),(12) reduce to the standard $\Lambda$CDM cosmology.
\begin{equation}
	\frac{{\dot H}}{{{H^2}}}  =  - \frac{3}{2}\Omega_{m}    - \frac{3}{2}\Omega_{DE} (1 + {\omega _{DE}}).
\end{equation}
 We consider $\omega_{m} = 0$ (The pressure of matter is zero).
 Furthermore, equation (13) gives 
immediately 
$\Omega_m=\Omega_
{m0} H_{0}
^2/a^3 H^2$ and recalling the fact that $\Omega_m + \Omega_{DE}=1$ we can 
obtain an expresssion for the Hubble parameter which reads as 
\be \label{h2}
H=\frac{\sqrt{\Omega_{m0}} H_{0}}{\sqrt{a^3 (1-\Omega_{DE})}}.
\ee
where a is scale factor.

 \subsection{Modified Friedmann equations  in presence non relativistic neutrino through kaniadakis entropy}

We start from modified Friedmann equation in presence of non relativistic neutrino in flat case $ k=0$,
\begin{eqnarray}
	\label{FR1}
	&H^2 = \frac{8\pi G}{3}({\rho _b} + {\rho _c} + {\rho _\nu } +\rho_{\rm r} + \rho_{DE})\\
	&\dot{H} = -4\pi G({\rho _b} + {\rho _c} + {\rho _\nu } +\rho_{\rm r} + {p_b} + {p_c} + {p_\nu } + p_{\rm r} \\ \nonumber&+\rho_{DE}+p_{DE}).
	\label{FR2}
\end{eqnarray}
Moreover, we add the radiation density $\rho_{\rm r}$ in above equations.

\begin{equation}
		\Omega_{m}  = \frac{{{\rho _b}}}{{3{H^2}}},\Omega_{\nu}  = \frac{{{\rho _\nu }}}{{3{H^2}}},\Omega_{c}  = \frac{{{\rho _c}}}{{3{H^2}}},\Omega_{r}  = \frac{{{\rho _r}}}{{3{H^2}}},\Omega_{DE}  = \frac{{{\rho _{DE}}}}{{3{H^2}}}.
\end{equation}
	were $ \Omega_{\rm b} $ is baryon density, $ \Omega_{\rm r} $ is radiation density, $ \Omega_{\rm c} $ is cold dark matter density, $ \Omega_{\nu} $ is neutrino density and $ \Omega_{DE} $  denotes as dark energy density.
Now, we can derive the following dynamical system:
 \begin{align}
 	\begin{split}
 		&\Omega_{m}^{'} =  - 3\Omega_{m}  - 2\Omega_{m} \frac{{\dot H}}{{{H^2}}},\\
 		&\Omega_{\nu} ^{'} =  - 3\Omega_{\nu} (1 + {\omega _\nu(z) }) - 2\Omega_{\nu} \frac{{\dot H}}{{{H^2}}},\\
 		&\Omega_{c}^{'} =  - 3\Omega_{c}  - 2\Omega_{c} \frac{{\dot H}}{{{H^2}}},\\
 		&\Omega_{DE}^{'} =  - 3\Omega_{DE} (1 + {\omega _{DE}}) - 2\Omega_{DE} \frac{{\dot H}}{{{H^2}}}\\
 &\omega_{\nu} ^{'}=\frac{2\omega_{\nu}}{z_{\rm dur}}(3\omega_{\nu}-1).
 	\end{split}	
 \end{align}

We shall
use the following ansatz for $\omega_{\nu}(z)$ \cite{Wali}
\begin{equation}
	\omega_{\nu}(z)=\frac{p_{\nu}}{\rho_{\nu}}=\left(1+tanh(\frac{ln(1+z)-z_{\rm eq}}{z_{\rm dur}})\right),
\end{equation}

where $z_{eq}$ determines the transition redshift where matter and radiation energy densities become equal and $z_{\rm dur}$ represents  how fast this transition is realized.
The  Friedmann constraint is:

 \begin{align}
 	\Omega_{r}  = 1 - \Omega_{m}  - \Omega_{\nu}  - \Omega_{c}  - \Omega_{DE}.
 \end{align}
 It is clear that in the case where $k=0$, the generalized Friedmann equations
 (17),(18) reduce to the standard $\Lambda$CDM cosmology. Moreover, we can ontain
 \begin{equation}
\frac{{\dot H}}{{{H^2}}}  =  - \frac{3}{2}\Omega_{\nu} (1 + {\omega _\nu }) - \frac{1}{2}\Omega_{r} (3 + {\omega _r}) - \frac{3}{2}\Omega_{DE} (1 + {\omega _{DE}}).
 \end{equation}
The above parameter is very useful for the connection between the theoretical model and observations. 
To investigate the  Hubble and $ S_{8} $  Tensions in the dynamical system model, we use the luminosity distance relation for pantheon and cc data (Direct method). We start from following $d_{L}$ relation.
\begin{align} \label{dl}
	d_{L}=(1+z)\int\frac{dz}{H(z)}.
\end{align}
By introducing the new variables $x_{d}=d_{L}$ and $x_{h}=H$ and
(Since $1+z\equiv\frac{1}{a}$, then $(1+z)\equiv e^{-N}$, $dz\equiv-e^{-N}dN$ and $dN\equiv Hdt$), the relation (24) can be converted to couple ODE differential equations
as
\begin{align}\label{dl3}
	&dx_{d}^{'}=-x_{d}-\frac{e^{-2N}}{x_{h}}\\
	&dx_{h}^{'}=\frac{{\dot H}}{{{H^2}}}  .x_{h}.\label{dl4}
\end{align}

\section{Numerical Analysis}

To analyze the data and extract the constraints on these cosmological parameters, we
used our modiﬁed version of the publicly available Monte Carlo Markov Chain package
CosmoMC \cite{Lewis}. This is equipped with a convergence diagnostic based on the Gelman and
Rubin statistic \cite{Gelman}, assuming $R - 1 < 0.02$, and implements an efﬁcient sampling of the pos-
terior distribution using the fast/slow parameter decorrelations \cite{Lewis1}. CosmoMC includes
support for the 2018 Planck data release \cite{43}. Moreover, we  used  CAMB code for anisotropy. The CAMB (Code for Anisotropies in the Microwave Background) is a software package for calculating the cosmic microwave background (CMB) and matter power spectra, as well as other cosmological observables. It is widely used in the cosmology community for theoretical predictions and analysis of observational data. Also, we used the Akaike Information Criteria (AIC).
The Akaike Information Criterion (AIC) is a statistical measure used to compare different statistical models based on their ability to fit the data while balancing the complexity of the model. The  AIC equation is:

\begin{equation}\label{key}
	AIC = \chi _{\min }^2 + 2\gamma	
\end{equation}
In these equations$\chi _{\min }^2  $ is the minimum value of $ {\chi ^2} $, $\gamma$ is the number of parameters of the given model.
All observational data where used in this paper are:\\
$\bullet$ Pantheon catalog:
We used updated the Pantheon + Analysis catalog consisting of 1071 SNe Ia covering the redshift range $0.001 < z < 2.3$\cite{42}.\\
$\bullet$ {CMB data}:
We used the latest large-scale cosmic microwave background (CMB) temperature and
polarization angular power spectra from the final release of Planck 2018 plikTTTEEE+lowl+lowE
\cite{43}. \\
$\bullet$ {BAO data}:
We also used the various measurements of the Baryon Acoustic Oscillations (BAO) from
different galaxy surveys \cite{43}, i.e.
6dFGS.(2011)\cite{44}, SDSS-MGS
\cite{45}.\\
$\bullet$ {CC data}: The 32 $H(z)$ measurements listed in Table I have a redshift range of $0.07 \leq z \leq 1.965$. The covariance matrix of the 15 correlated measurements originally from Refs. (\cite{Moresco}; \cite{Moresco1}; \cite{Moresco2}) , discussed in Ref. \cite{Moresco3}, can be found at https://gitlab.com/mmoresco/CCcovariance/.

\begin{table}[t!]
	\centering
	\caption{32 $H(z) 
		$  data.}\label{tab:hz}
	\setlength{\tabcolsep}{7.5mm}{
		\begin{tabular}{lcc}
			\hline
			$z$ & $H(z) (km/s/Mpc)$ & $\sigma$\\
			\hline
			0.07 & $69.0$ & 19.6\\
			0.09 & $69.0$ & 12.0\\
			0.12 & $68.6$ & 26.2\\
			0.17 & $83.0$ & 8.0\\
			0.2 & $72.9$ & 29.6\\
			0.27 & $77.0$ & 14.0\\
			0.28 & $88.8$ & 36.6\\
			0.4 & $95.0$ & 17.0\\
			0.47 & $89.0$ & 50.0\\
			0.48 & $97.0$ & 62.0\\
			0.75 & $98.8$ & 33.6\\
			0.88 & $90.0$ & 40.0\\
			0.9 & $117.0$ & 23.0\\
			1.3 & $168.0$ & 17.0\\
			1.43 & $177.0$ & 18.0\\
			1.53 & $140.0$ & 14.0\\
			1.75 & $202.0$ & 40.0\\
			0.1791 & 74.91 & 4.00\\
			0.1993 & 74.96 & 5.00\\
			0.3519 & 82.78 & 14\\
			0.3802 & 83.0 &  13.5\\
			0.4004 & 76.97 &  10.2\\
			0.4247 & 87.08 &  11.2\\
			0.4497 & 92.78 &  12.9\\
			0.4783 & 80.91 &  9\\
			0.5929 & 103.8 & 13\\
			0.6797 & 91.6 & 8\\
			0.7812 & 104.5 & 12\\
			0.8754 & 125.1 & 17\\
			1.037 & 153.7 & 20\\
			1.363 & 160.0 & 33.6\\
			1.965 & 186.5 & 50.4\\
			\hline
	\end{tabular}}
\end{table}

By using  the first and the second scenario, we can  put constraints on the following cosmological parameters: the Baryon energy density ${\Omega _b}{h^2}$, the cold dark matter energy density $\Omega_{c}h^{2}$, the neutrino density ${\Omega _\nu}$, the Kaniadakis parameter k, the ratio of the sound horizon at decoupling to the angular diameter distance to last scattering $\theta_{MC}$, the optical depth to reionization $\tau$, the amplitude and the spectral index of the primordial scalar perturbations $A_{s}$ and $n_{s}$. The results  obtained for two scenarios are in Tables 5 and 8.

 \begin{table}
 	\begin{center}
 		 	\caption{Flat priors for the cosmological parameters.}
 		 		\resizebox{0.3\textwidth}{!}{
 		\begin{tabular}{|c|c|}
 				\hline 
 			Parameter                    & Prior\\
 			\hline 
 			$\Omega_{b} h^2$             & $[0.005,0.1]$\\
 			$\Omega_{c} h^2$             & $[0.005,0.1]$\\
 			$\tau$                       & $[0.01,0.8]$\\
 			$n_s$                        & $[0.8,1.2]$\\
 			$\log[10^{10}A_{s}]$         & $[1.6,3.9]$\\
 			$100\theta_{MC}$             & $[0.5,10]$\\ 
 			$k$                          & $(-1,1)$\\ 
 			${\Omega _\nu}$              & $[0.001,0.005]$\\
 			\hline 
 		\end{tabular}
 	}
 	\end{center}
 	\label{tab:priors}
 \end{table}
 
\section{kaniadakis horizon  entropy without neutrinos}
The Kaniadakis horizon entropy without neutrinos is a theoretical model that aims to provide an explanation for the accelerated expansion of the universe without the use of dark energy. The model utilizes the Kaniadakis entropy, which is a generalization of the Bekenstein-Hawking entropy and can be applied to a broader range of physical systems.

To derive the Kaniadakis horizon entropy without neutrinos, it is assumed that the entropy associated with the apparent horizon of the Friedmann-Robertson-Walker (FRW) Universe follows the Kaniadakis prescription. This prescription is a generalization of the standard Boltzmann-Gibbs entropy and allows for a non-extensive distribution of matter and energy.

The resulting modified Friedmann equations are similar to the standard Friedmann equations but contain an additional term that can be interpreted as an effective dark energy term. This term is proportional to the Kaniadakis parameter K, which determines the deviation from conventional statistical behavior.

The Kaniadakis parameter K usually ranges between -1 and 1. When K is equal to zero, the modified Friedmann equations reduce to the standard Friedmann equations, which do not account for dark energy. However, as the value of K increases, the effective dark energy density increases and the universe accelerates at a faster rate.
The results obtained for different combinations of datasets are as follows:

$\bullet$ For CMB + Pantheon data:
We found $ H_{0} = 69.52 \pm 1.87$  $kms^{-1} Mpc^{-1}$ at 68\% CL, which is close to Planck 2018 results and there is a 1.09$ \sigma $  with Planck result and 1.92$ \sigma $ tension with R22. Also, the value obtained for $ S_{8} $  is $ 0.809\pm 0.059$. The tension of this result with the DES, and kiDS,  and, Planck are: 0.38$ \sigma $, 0.53$ \sigma $, and 0.71$ \sigma $, respectively.
 
$\bullet$ For  CMB + CC data:
We found $ H_{0} = 69.03 \pm 1.23 $  $kms^{-1} Mpc^{-1}$ at 68\% CL,  which is close to Planck 2018 results and there is a 1.22$ \sigma $  with Planck result and 2.66$ \sigma $ tension with R22. Also, the value obtained for $ S_{8} $  is $ 0.801\pm 0.041 $. This result is consistent with the Planck results at 0.72$ \sigma $ tension, and 0.56$ \sigma $  with  DES,  and 0.82$ \sigma $  with kiDS. 

$\bullet$ For CMB + BAO data:
We found $ H_{0} = 68.81 \pm 1.33 $  $kms^{-1} Mpc^{-1}$ at 68\% CL, which is close to Planck 2018 results and there is a 0.99$ \sigma $  with Planck result  and 2.68$ \sigma $ tension with R22. Also, the value obtained for $ S_{8} $  is $ 0.81 \pm 0.12 $. This result is in complete agreement with Planck's result at 0.18$ \sigma $ tension and  with the DES at 0.28$ \sigma $. Also, this result is in the 0.36$ \sigma $  with kiDS.
  
$\bullet$ For CMB + BAO + Pantheon + CC data:
We found $ H_{0} = 69.07 \pm 1.51 $  $kms^{-1} Mpc^{-1}$ at 68\% CL,. which is close to Planck 2018 results and there is a 1.04$ \sigma $  with Planck result and 2.39$ \sigma $ tension with R22 result. Also, the value obtained for $ S_{8} $  is $ 0.802\pm 0.043 $. This result is in complete agreement with Planck's result at 0.66$ \sigma $ tension and  with the DES at 0.56$ \sigma $. Also, this result is in the 0.81$ \sigma $  with kiDS.\\
\\
All results are in table 3, 4.
The results obtained in the first scenario are shown in Figures 1,2, and 4. Also, Fig.3 show the comparison  results of the  $S_{8} $ , $\Omega_b h^2   $, $\Omega_c h^2   $, $n_s  $, ${\rm{ln}}(10^{10} A_s)$, $100\theta_{MC} $ for different combination dataset for kaniadakis entropy  without neutrinos.
All results are in table 5.

	Figure 1 illustrates the comparison results of the Hubble constant, denoted as $H_{0} = 69.07 \pm 1.51$ km/s/Mpc, obtained from the combined dataset (CMB + BAO + Pantheon + CC) utilizing Kaniadakis entropy without neutrinos. The error margin is provided at a 68$\%$ confidence level.
	
	This specific combination of cosmological datasets aims to offer a comprehensive understanding of the Hubble constant, incorporating contributions from cosmic microwave background (CMB), baryon acoustic oscillations (BAO), Pantheon supernova data, and cosmic chronometers (CC).
	
	The presented results showcase not only the central value of the Hubble constant but also the associated uncertainty, allowing for a more robust interpretation of the cosmological implications. The incorporation of Kaniadakis entropy without neutrinos further adds a nuanced perspective to the analysis.
Figure 2 denotes the Comparison  results of the  $\Omega_{m}$ , $S_{8}$ according to $H_{0}$ for different combination dataset for kaniadakis entropy  without neutrinos.  This analysis undertakes a comparative examination of the cosmological parameters, $\Omega_{m}$ and $S_{8}$, with respect to the Hubble constant ($H_{0}$) within the framework of different combination datasets. The study employs Kaniadakis entropy without neutrinos and considers key cosmological datasets, including cosmic microwave background (CMB), baryon acoustic oscillations (BAO), Pantheon supernova data, and cosmic chronometers (CC). By scrutinizing the dependencies and variations of $\Omega_{m}$ and $S_{8}$ in response to the chosen $H_{0}$ values, this analysis elucidates the interconnected nature of these crucial cosmological parameters.
Also, figure 3, presents a comparative analysis of cosmological parameters, including $S_{8}$, $\Omega_b h^2$, $\Omega_c h^2$, $n_s$, ${\rm{ln}}(10^{10} A_s)$,$\tau$ and $100\theta_{MC}$, across various combination datasets. The study employs Kaniadakis entropy without neutrinos. Each parameter holds significance in characterizing different aspects of the universe.
Figure 4 demonstrate the comparison of $H_{0}$ measurement for different combination of data sets with results of Planck 2018 and R22 for kaniadakis entropy  without neutrinos.

	\begin{table*}[h]
		\centering
		\caption{Comparison  $H_0$, $H_0$ Tension for different combinations of data for Kaniadakis without neutrinos}
		\begin{tabular}{|l|c|c|c|c}
			\hline
			dataset& $H_{0}$ ($\mathrm{km \, s^{-1} \, Mpc^{-1}}$) & Tension with Planck & Tension with R22\\
			\hline
			CMB  & $68.79\pm 1.27$ & $0.32\sigma$ & $2.75\sigma$ \\
			CMB + Pantheon & $69.52\pm 1.87$ & $1.09\sigma$ & $1.92\sigma$ \\
			CMB + CC & $69.03\pm 1.23$ & $1.22\sigma$ & $2.66\sigma$  \\
			CMB + BAO & $68.81\pm 1.33$ & $0.99\sigma$ & $2.68\sigma$  \\
			CMB + BAO + Pantheon + CC & $69.07\pm 1.51$ & $1.04\sigma$ & $2.39\sigma$ \\
			\hline
		\end{tabular}
	\end{table*}
	
	\begin{table*}[h]
		\centering
		\caption{Comparison  $S_8$, and  $S_8$ Tension for different combinations of data for Kaniadakis without neutrinos}
		\begin{tabular}{|l|c|c|c|c|}
			\hline
		dataset	& $S_{8}$ & Tension with DES & Tension with KiDS & Tension with Planck \\
			\hline
			CMB  & $0.816 \pm 0.086$ & $0.45\sigma$ & $0.57\sigma$ & $0.18\sigma$ \\
			CMB + Pantheon & $0.809 \pm 0.059$ & $0.38\sigma$ & $0.53\sigma$ & $0.38\sigma$ \\
			CMB + CC & $0.801 \pm 0.041$ & $0.56\sigma$ & $0.82\sigma$ & $0.72\sigma$ \\
			CMB + BAO & $0.81 \pm 0.12$ & $0.28\sigma$ & $0.36\sigma$ & $0.18\sigma$ \\
			CMB + BAO + Pantheon + CC & $0.802 \pm 0.043$ & $0.56\sigma$ & $0.81\sigma$ & $0.66\sigma$ \\
			\hline
		\end{tabular}
	\end{table*}

	\begin{table}[h]
		\centering
		\caption{Cosmological Parameter Results for Different Datasets for Kaniadakis horizon entropy without neutrinos}
		\resizebox{0.9\textwidth}{!}{	
		\begin{tabular}{|l|c|c|c|c|c|}
			\hline
				Parameter & CMB+BAO & CMB+CC &CMB+Pantheon & CMB+ALL& CMB \\
			\hline
			${\Omega_b h^2}$ & $0.02234 \pm 0.0002$ & $0.02223 \pm 0.00028$ & $0.02226 \pm 0.00023$ & $0.02229 \pm 0.00019$ &$0.02233 \pm 0.00027$\\
			\hline
			${\Omega_c h^2}$ & $0.1178 \pm 0.0025$ & $0.1200 \pm 0.0037$ & $0.1200 \pm 0.0037$ & $0.1181 \pm 0.0025$&$0.1176 \pm 0.0028$ \\
			\hline
			${100\theta_{MC}}$ & $1.04119 \pm 0.00039$ & $1.04089 \pm 0.00057$ & $1.04089 \pm 0.00057$ & $1.04105 \pm 0.00040$ &$1.04120 \pm 0.00037$\\
			\hline
			${\tau}$ & $0.0559^{+0.0054}_{-0.0078}$ & $0.0544 \pm 0.0076$ & $0.0553^{+0.0054}_{-0.0080}$ & $0.0554^{+0.0051}_{-0.0076}$ &$0.0558^{+0.0057}_{-0.0079}$\\
			\hline
			${{\rm{ln}}(10^{10} A_s)}$ & $3.042^{+0.013}_{-0.016}$ & $3.044 \pm 0.018$ & $3.046^{+0.015}_{-0.018}$ & $3.041^{+0.013}_{-0.016}$&$3.041^{+0.014}_{-0.018}$ \\
			\hline
			${n_s}$ & $0.9635 \pm 0.0062$ & $0.9677 \pm 0.0087$ & $0.9679 \pm 0.0087$ & $0.9648 \pm 0.0064$&$0.9632 \pm 0.0065$ \\
			\hline
		\end{tabular}
	}
	\end{table}

\begin{figure}
	\includegraphics[width=9 cm]{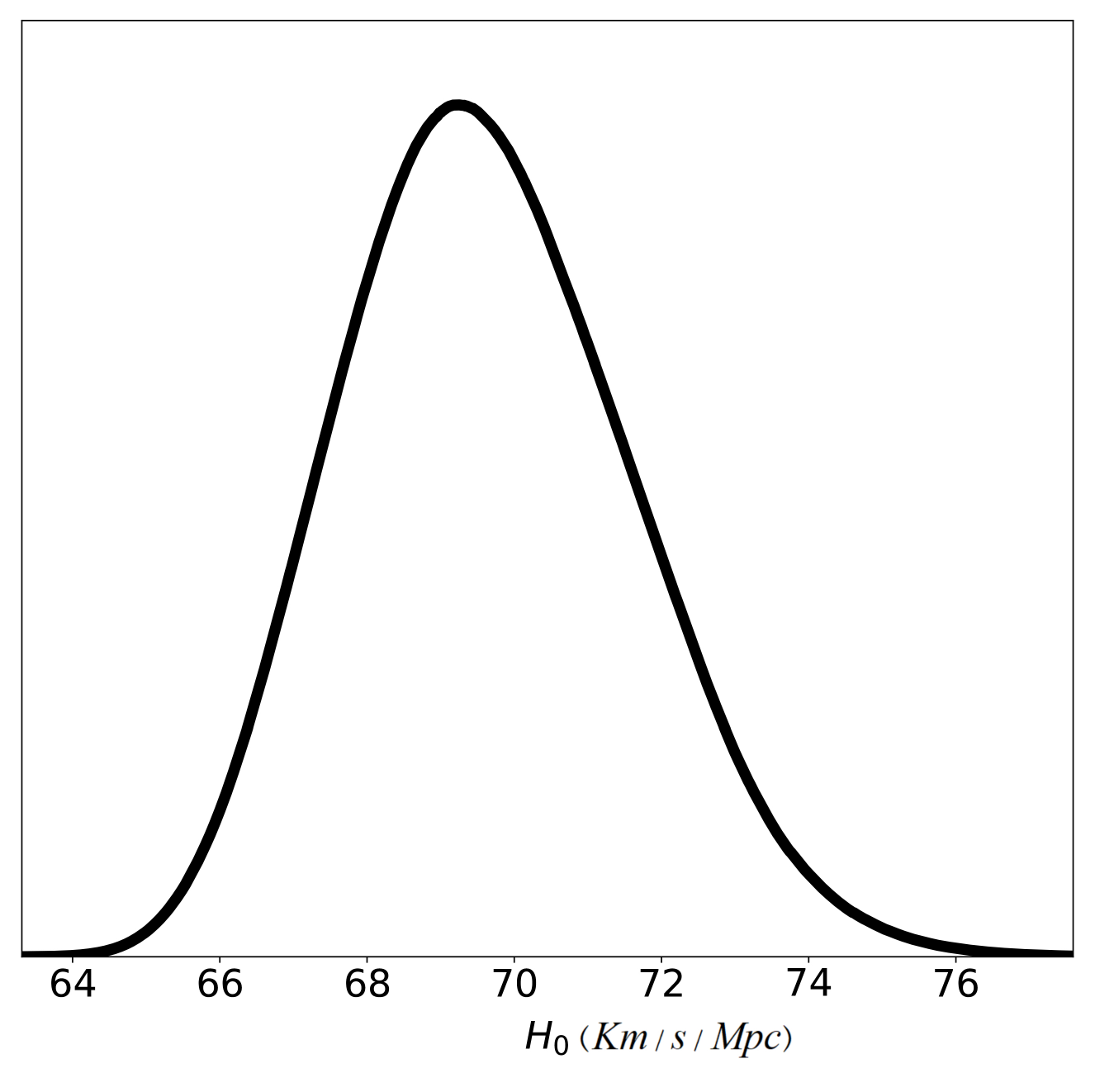}
	\vspace{-0.12cm}
	\caption{\small{Comparison  results of the  $H_{0}$ for  combination dataset(CMB + BAO + Pantheon + CC) for kaniadakis entropy  without neutrinos. }}\label{fig:omegam2}
\end{figure}
\begin{figure}
	\includegraphics[width=8.5 cm]{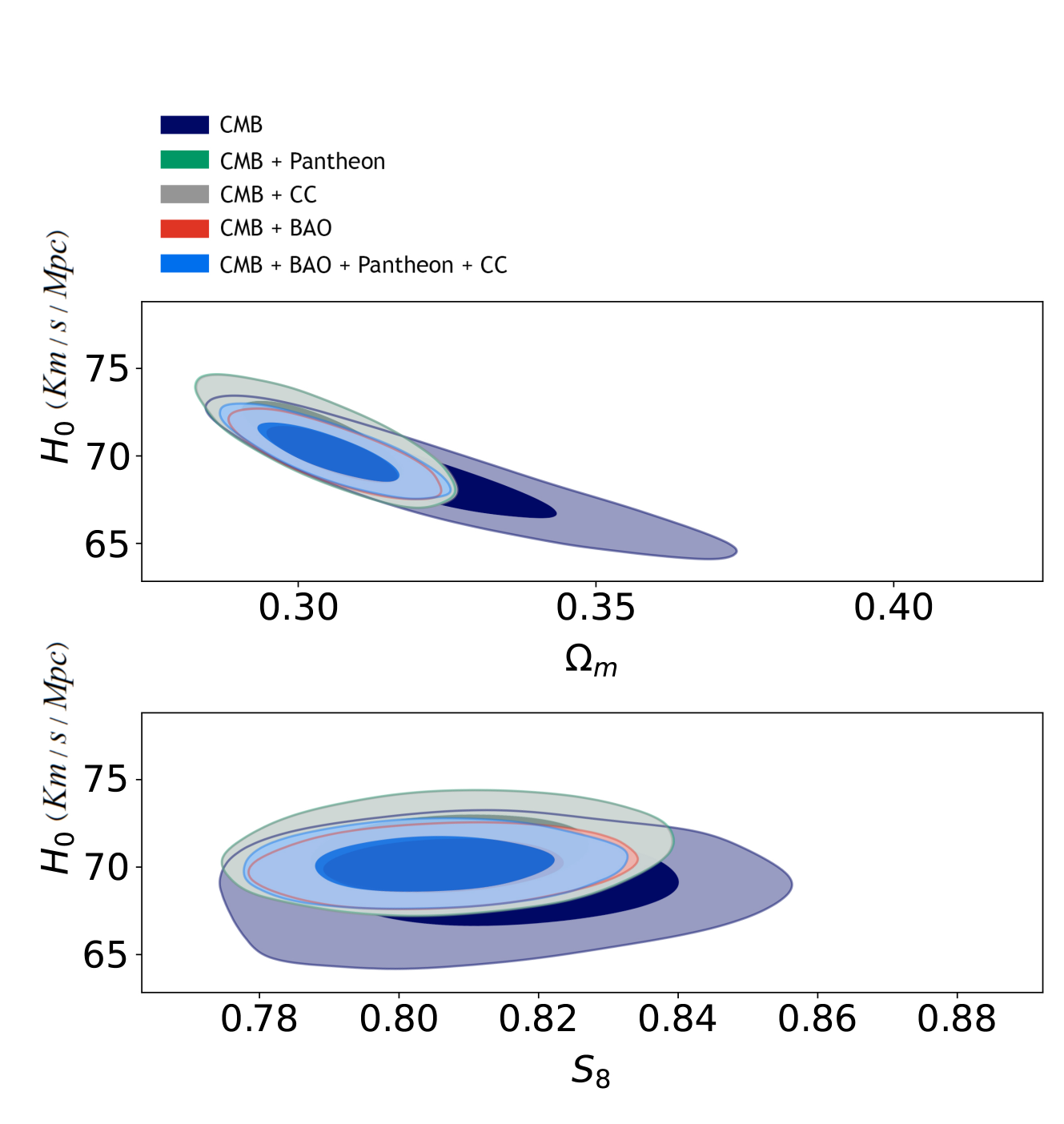}
	\vspace{-0.12cm}
	\caption{\small{Comparison  results of the  $\Omega_{m}$ , $S_{8}$ according to $H_{0}$ for different combination dataset for kaniadakis entropy  without neutrinos. }}\label{fig:omegam2}
\end{figure}
\begin{figure*}
	\includegraphics[width=16.5 cm,height=18 cm]{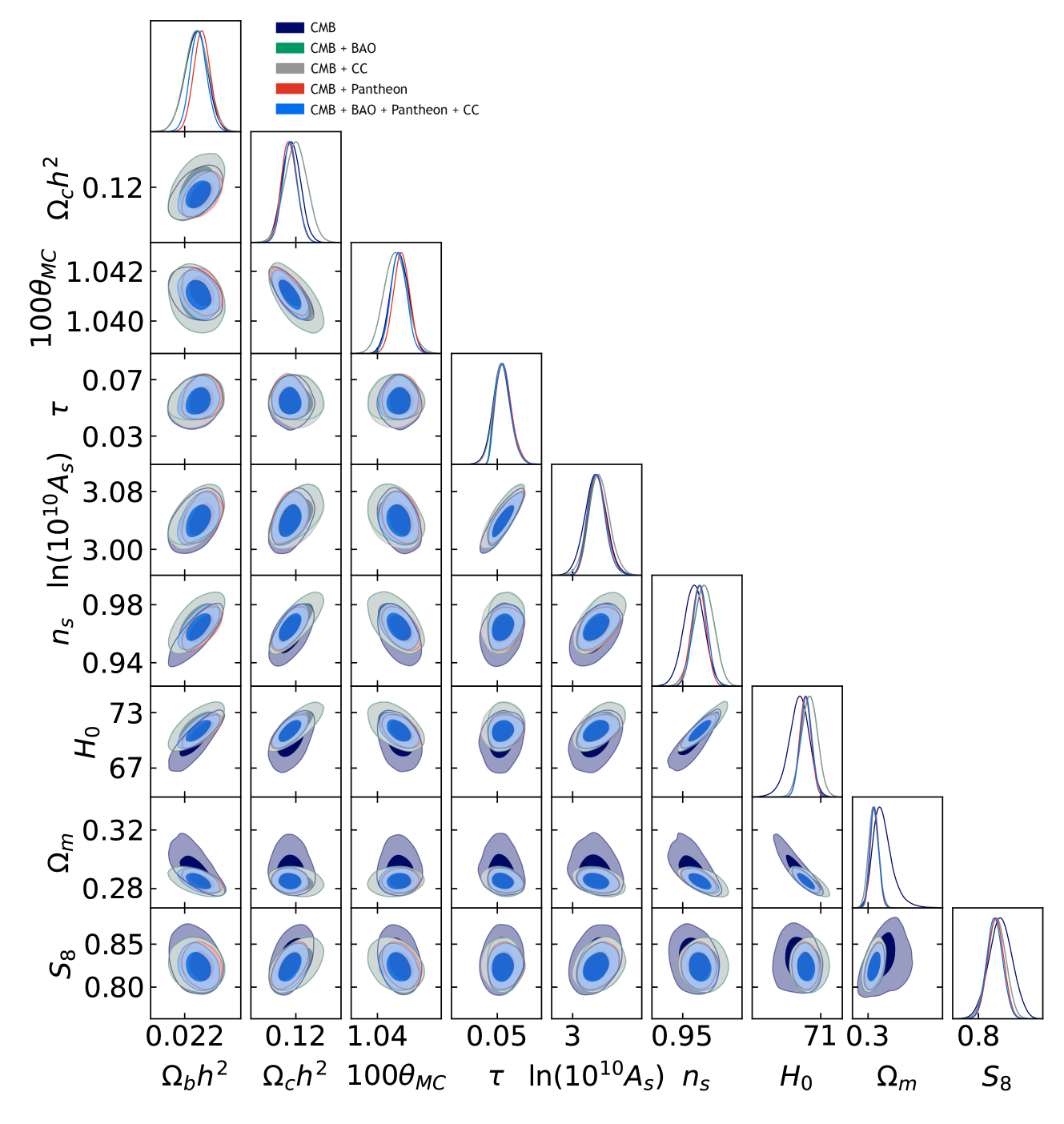}
	\vspace{-0.12cm}
	\caption{\small{Comparison  results of the  $S_{8} $ , $\Omega_b h^2   $, $\Omega_c h^2   $, $n_s  $, ${\rm{ln}}(10^{10} A_s)$, $100\theta_{MC} $, $H_{0}(km/s/Mpc
			)$ for different combination dataset for kaniadakis entropy  without neutrinos.}}\label{fig:omegam2}
\end{figure*}
\begin{figure*}
	\includegraphics[width=16.5 cm,height=10 cm]{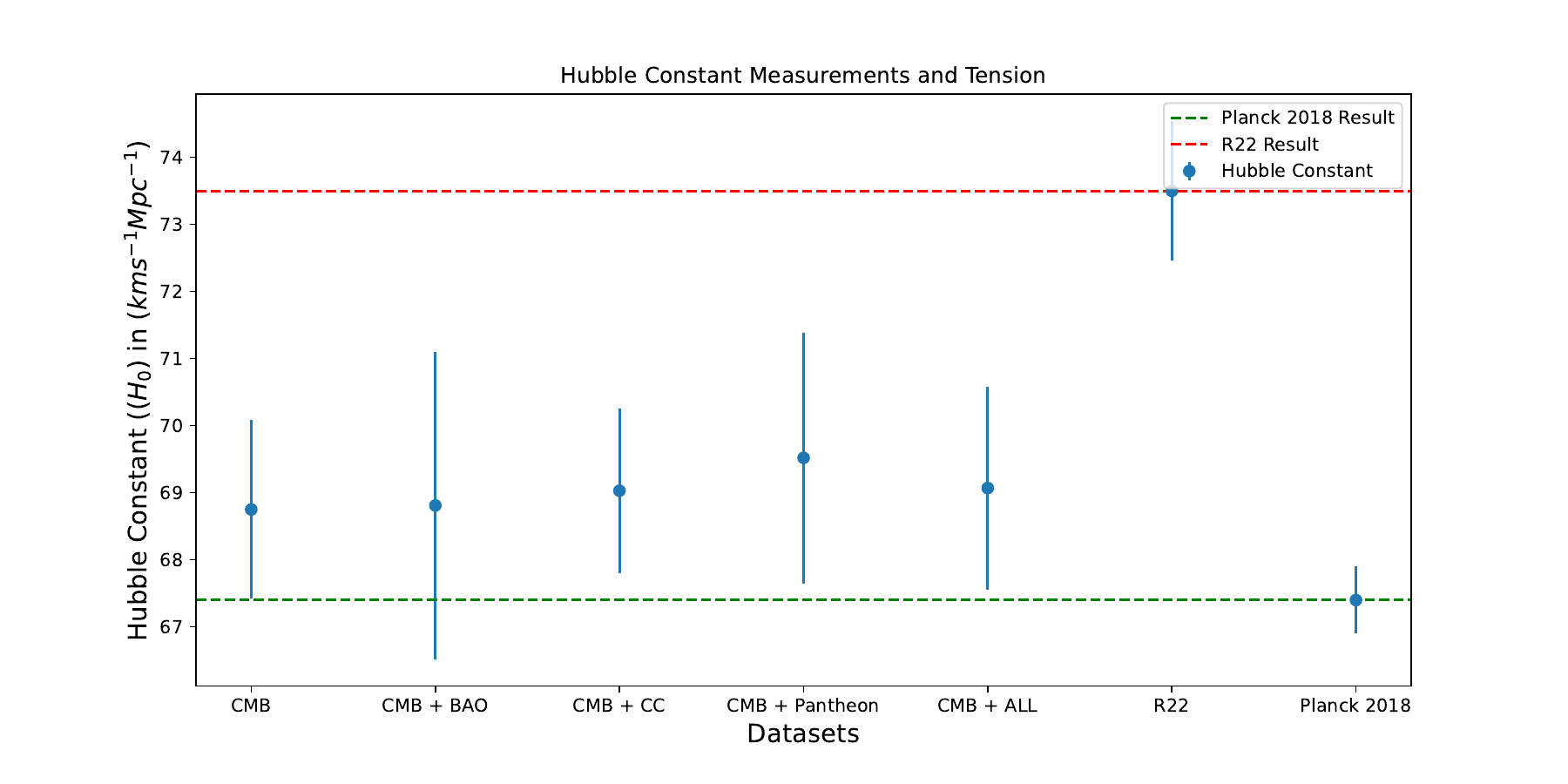}
	\vspace{-0.12cm}
	\caption{\small{The comparison of $H_{0}$ measurement for different combination of data sets with results of Planck 2018 and R22 for kaniadakis entropy  without neutrinos.}}\label{fig:omegam2}
\end{figure*}

The obtained best fit result, indicating $\Lambda = 0.0052$, offers intriguing insights into the role of the cosmological constant and the k parameter within the context of the Kaniadakis horizon entropy. This relatively low value of $\Lambda$ suggests that the cosmological constant, often associated with dark energy, exerts minimal influence on the observed accelerated expansion of the universe in this particular model.

The key implication of this finding is that the k parameter emerges as a potential substitute for dark energy in explaining the accelerated expansion. The k parameter, integral to the Kaniadakis entropy model, seems to play a significant role in accounting for the observed cosmological phenomena. This suggests a departure from traditional dark energy formulations, providing an alternative avenue for understanding the dynamics of the universe.

Furthermore, the conclusion drawn from this analysis posits that the Kaniadakis horizon entropy, incorporating the k parameter, possesses the explanatory power to account for the accelerated expansion without the explicit need for dark energy. This challenges conventional paradigms and prompts a reevaluation of the fundamental constituents driving the observed cosmic dynamics.

In summary, the identified best fit value for $\Lambda$ and the suggested substitution of the k parameter for dark energy underscore the potential of the Kaniadakis horizon entropy model to offer a distinct and compelling explanation for the accelerated expansion of the universe. This not only expands our theoretical understanding of cosmic evolution but also paves the way for further exploration into novel approaches to cosmological modeling.

\section{kaniadakis entropy in presence of neutrino}
In the early universe, neutrinos were relativistic, meaning they traveled at speeds close to the speed of light and did not possess any measurable mass. This relativistic nature made them influential players in the dynamics of the cosmos during its early stages.
As the universe evolved, neutrinos underwent a transition from being relativistic to becoming non-relativistic. The total mass of neutrinos, denoted as $ \sum {m_\nu} $, becomes a significant parameter in cosmological studies.

	\section*{Transition from Relativistic to Non-Relativistic Phase for Neutrinos}
	
	The transition from the relativistic to the non-relativistic phase for neutrinos marks a crucial epoch in the evolution of the early universe. Neutrinos, being nearly massless and electrically neutral, exhibit relativistic behavior during the high-temperature and high-energy conditions of the early universe. As the universe expands and cools, neutrinos undergo a significant transition, impacting their dynamics and contributions to cosmic evolution.
	
	During the relativistic phase, neutrinos travel at speeds close to the speed of light, and their behavior is described by relativistic equations. At this stage, their mass is considered negligible, and they interact primarily through weak force interactions. As the universe expands, temperatures decrease, and neutrinos eventually enter a phase where their mass becomes non-negligible, marking the onset of the non-relativistic regime.	
 Neutrinos, which were initially considered massless during the relativistic phase, start to exhibit non-negligible mass effects. This transition is essential for understanding their impact on the large-scale structure of the universe\cite{52}. Relativistic neutrinos are known for their free-streaming behavior, meaning they travel long distances without significant interactions. As they become non-relativistic, their free-streaming behavior diminishes, leading to increased clustering and gravitational interactions\cite{53}. The transition from relativistic to non-relativistic neutrinos has implications for structure formation in the universe. In the relativistic phase, the free-streaming of neutrinos suppresses the growth of cosmic structures on small scales. As they become non-relativistic, their clustering enhances, influencing the formation of cosmic structures like galaxies and galaxy clusters. The transition affects the cosmic microwave background anisotropies. Relativistic neutrinos contribute to the radiation content of the early universe and influence the CMB. Their transition to non-relativistic speeds alters their contribution to the energy density, influencing the CMB power spectrum.
As we can see in Fig.5, considering the mass of neutrinos lead to several changes in the Cosmic Microwave Background (CMB) power spectrum compared to the standard $\Lambda $CDM  model and shifted the peaks.  Additionally, their effects on the growth of large-scale structure contribute to the Hubble tension by influencing measurements of the Hubble constant at different cosmic epochs. 
	\begin{figure*}
		\includegraphics[width=17.5 cm]{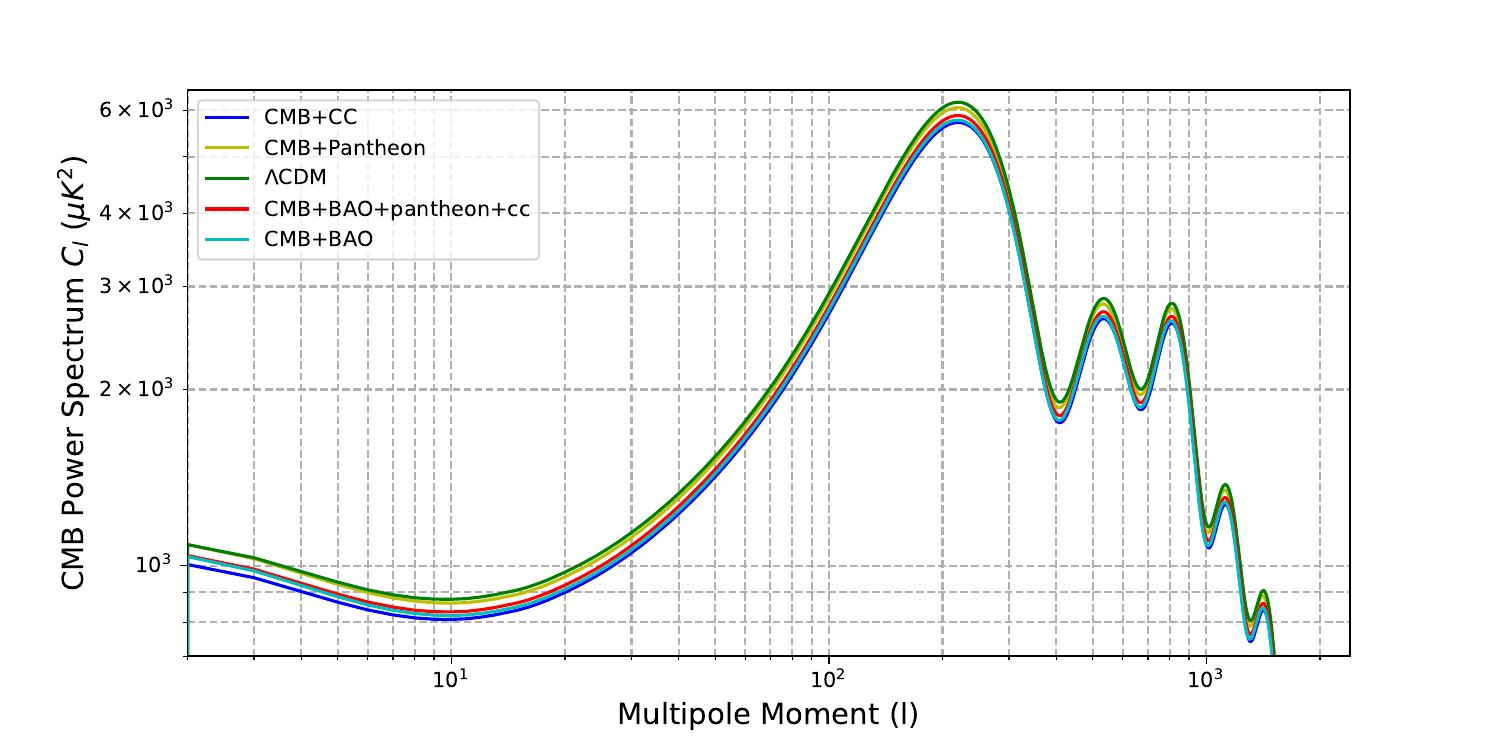}
		\vspace{-0.12cm}
		\caption{\small{Comparison the CMB power spectrum   combination dataset for kaniadakis entropy  in presence of neutrinos with $\Lambda $CDM model. }}\label{fig:omegam2}
	\end{figure*}

	\subsection*{Neutrino Transition and its Cosmological Implications}
	
	The transition from the relativistic to the non-relativistic phase for neutrinos is intricately linked to key cosmological parameters, notably the Hubble constant (\(H_0\)) and the parameter \(S_8\), which characterizes the amplitude of matter fluctuations in the universe.
	
$\bullet$\textbf{Hubble Constant (\(H_0\))}
 During the relativistic phase, neutrinos contribute significantly to the energy density of the early universe, influencing the cosmic dynamics and expansion rate.
The transition from relativistic to non-relativistic phases alters the energy density and dynamics of neutrinos. This transition is pertinent to the Hubble tension, where variations in \(H_0\) measurements from different observational methods are observed.
 Neutrinos, especially during their relativistic phase, contribute to the overall energy density of the universe. Understanding their behavior during the transition is crucial for precise modeling of the components influencing \(H_0\) measurements.

$\bullet$\textbf{\(S_8\) Tension}
 The transition of neutrinos from relativistic to non-relativistic phases affects their role in the formation of cosmic structures. Relativistic neutrinos exhibit free-streaming behavior, influencing the growth of structures on small scales.
 The non-relativistic phase allows neutrinos to cluster more, impacting the matter distribution in the universe. This clustering behavior is relevant to the \(S_8\) parameter, which characterizes the amplitude of matter fluctuations.
 The behavior of neutrinos during the transition contributes to tensions in \(S_8\), particularly if different observational methods yield varying estimates of this parameter. A comprehensive understanding of neutrino dynamics is essential for addressing tensions and refining our knowledge of cosmic evolution.
 A constraint on the total mass of neutrinos can be established using the relation $ \Omega_\nu =  \frac{{\sum {m_\nu}}}{{93.14h^2}} $, where $ h $ is the reduced Hubble constant. This constraint provides a link between the cosmological parameters $ (h, \Omega_\nu) $ and the total neutrino mass. Hence if the parameters
(h, $\Omega_{\nu}$ ) are constrained then the parameter $ \sum {m_\nu }$ is constrained automatically. Then we investigate the effect of adding the neutrino term to the density of matter to reduce the Hubble Tension using the dynamical system method. After that we will calculate the value of $S_8$ by estimating the value of the $ \Omega {\rm m }$ in the present time. 
To understand the impact of neutrinos on the evolution of the universe, particularly in addressing the Hubble tension, the dynamical system method is employed. This method allows for the investigation of the dynamical behavior of the cosmological parameters over time. By adding the neutrino term to the density of matter, one aims to alleviate or explain the Hubble tension, which refers to discrepancies in the measurements of the Hubble constant from different observational methods. After incorporating the neutrino term and addressing the Hubble tension, the next step involves estimating the value of $ \Omega_m $ (density parameter for matter) in the present time. This estimation then allows for the calculation of $ S_8 $, a parameter that characterizes the amplitude of matter fluctuations in the universe.

From analysis, we find that
 $\sum m_{\nu}<0.276$eV \ \ (95$\% $CL.)  for CMB data,  $\sum m_{\nu}<0.113$eV \ \ (95$\% $CL.)  for CMB+BAO , and for (CMB + CC)we find $ \sum m_{\nu}<0.126$eV \ \ (95$\% $CL.),  and for (CMB + Pantheon) we find $ \sum m_{\nu}<0.146$eV \ \ (95$\% $CL.), and for combination of full data (CMB+BAO+CC+Pantheon) we find $ \sum m_{\nu}<0.116$eV \ \ (95$\% $CL.) which is fully agreement with \cite{43}. It seems that this is a very good model to estimate the mass of neutrinos because the results obtained from this model are in broad agreement with observation, and finally we can use it for other purposes. 
 Figure 6 shows the constraints at the (95$\% $CL.) two-dimensional contours for $\sum m_{\nu}$ in kaniadakis entropy  with neutrinos.
 \begin{figure}
 	\includegraphics[width=8.5 cm]{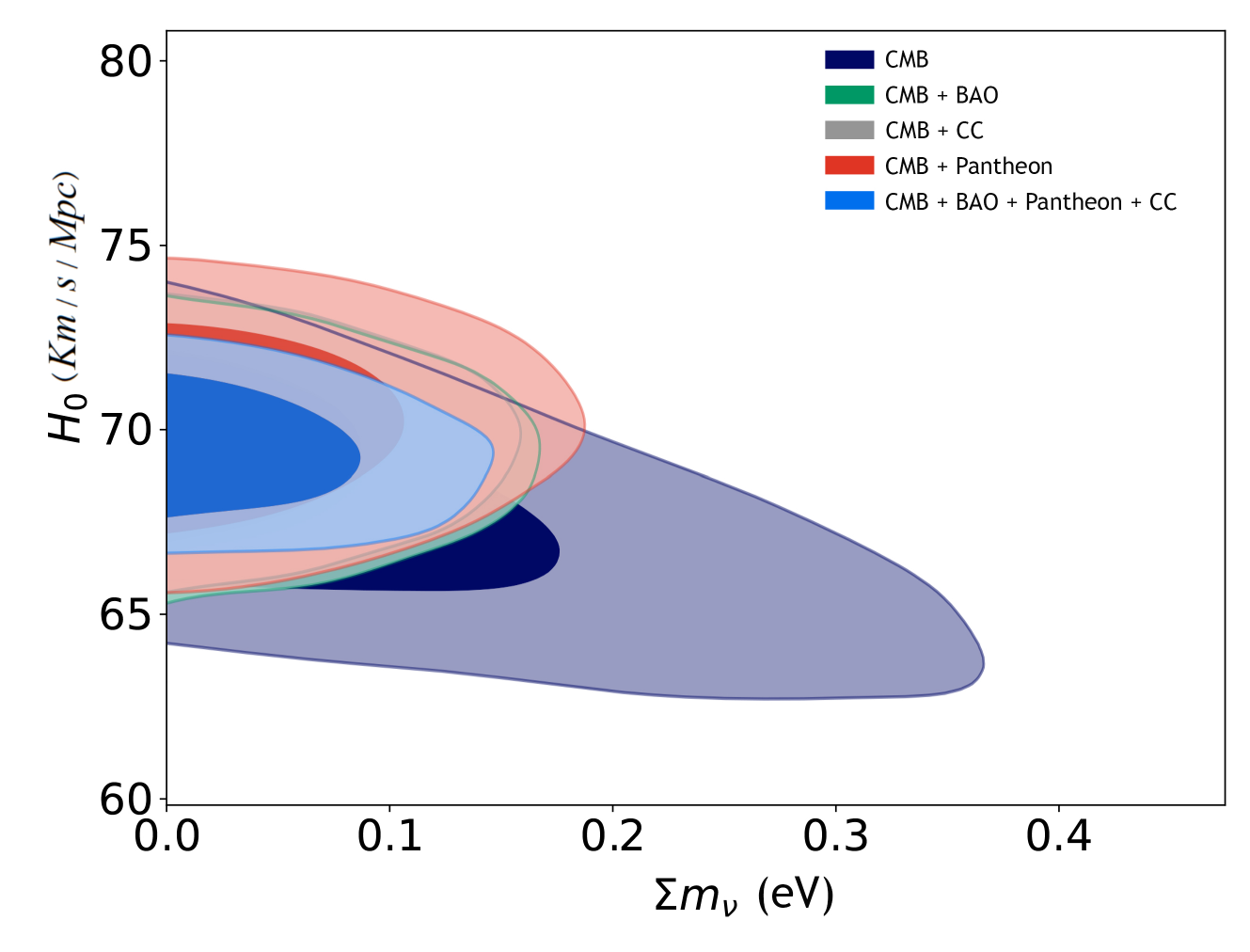}
 	\vspace{-0.12cm}
 	\caption{\small{Constraints at the (95$\% $CL.) two-dimensional contours for $\sum m_{\nu}$ in kaniadakis entropy  with neutrinos }}\label{fig:omegam2}
 \end{figure}

If we add the neutrino term, we find:\\
$\bullet$ For CMB + Pantheon data:
We found $ H_{0} = 71.68 \pm 2.2 $  $kms^{-1} Mpc^{-1}$ at 68\% CL, which is close to Planck 2018 results and there is a 1.89$ \sigma $ tension with Planck result  and 0.89$ \sigma $  with R22. Also, the value obtained for $ S_{8} $  is $ 0.788\pm 0.047 $. This result is consistent with the kiDS at 0.45$ \sigma $  and there is a 0.9$ \sigma $  with the Planck result. For comparing with DES, there is a 0.24$ \sigma $ at 68\% CL, i.e.\\

$\bullet$  For CMB + CC data:
We found $ H_{0} = 70.97 \pm 1.6 $  $kms^{-1} Mpc^{-1}$ at 68\% CL, which is close to Planck 2018 results and there is a 2.12$ \sigma $ tension with Planck result  and 1.43$ \sigma $ tension with R22. Also, the value obtained for $ S_{8} $  is $ 0.788\pm 0.031 $. This result is very close to the kiDS result(0.15$ \sigma $)  and there is a 1.3$ \sigma $ tension with the Planck result. Moreover, the tension with DES is 0.33$ \sigma $ which is in same as Planck 2018.\\

$\bullet$ For CMB + BAO data:
We found $ H_{0} = 71.56 \pm 1.51 $  $kms^{-1} Mpc^{-1}$ at 68\% CL, which is close to Planck 2018 results and there is a 2.61$ \sigma $ tension with Planck result  and 1.91$ \sigma $ tension with R22. The value obtained for $ S_{8} $  is $ 0.789\pm 0.049 $. This result is in broad with the kiDS(0.45$ \sigma $)  and there are  0.84$ \sigma $ and 0.25$ \sigma $  with Planck result, and DES, respectively.\\

$\bullet$ For CMB + BAO + Pantheon + CC data:
We found $ H_{0} = 70.61 \pm 1.49 $  $kms^{-1} Mpc^{-1}$ at 68\% CL, which is close to Planck 2018 results and there is a 2.04$ \sigma $ tension with Planck result  and 1.66$ \sigma $  with R22. Moreover, the value obtained for $ S_{8} $  is $ 0.788\pm 0.032 $. This result is fully in agreement with kiDS at 0.65$ \sigma $ tension. Also,  there is a 1.27$ \sigma $ with the Planck result and  0.33$ \sigma $  with DES. \\
All results are shown in tables 6,7.
The results obtained in the second scenario are shown in Figures 6,7, and 9. Also, Fig.8 show the comparison  results of the  $S_{8} $ , $\Omega_b h^2   $, $\Omega_c h^2   $, $n_s  $, $\tau$, ${\rm{ln}}(10^{10} A_s)$, $100\theta_{MC} $ for different combination dataset for kaniadakis entropy  in presence of neutrinos.
All results are in table 8.
In the context of the Figure 7 comparison, the investigation focuses on the Hubble constant within the framework of a combination dataset (CMB + BAO + Pantheon + CC) using Kaniadakis entropy with neutrinos. The juxtaposition of this dataset against the Planck and R22 results highlights the subtle intricacies and potential discrepancies in our understanding of the Hubble constant. 
The analysis reveals a measured value of the Hubble constant, $H_{0} = 70.61 \pm 1.49 , \mathrm{kms^{-1} Mpc^{-1}}$, at a 68\% confidence level. This result closely aligns with the Planck 2018 findings, indicating a good agreement within the uncertainties. Notably, there emerges a tension of 2.04$\sigma$ with the Planck result and 1.66$\sigma$ with the R22 outcome, emphasizing a noteworthy deviation from these established cosmological measurements.
Figure 8 demonstrate the comparison  results of the  $\Omega_{m}$ , $S_{8}$ according to $H_{0}$ for different combination dataset for kaniadakis entropy  with neutrinos. Moreover, Figure 9 illustrate the comparison  results of the  $S_{8} $ , $\Omega_b h^2   $, $\Omega_c h^2   $, $n_s  $, ${\rm{ln}}(10^{10} A_s)$,$\tau$, $100\theta_{MC} $ for different combination dataset for kaniadakis entropy  with neutrinos. 
Figure 10 demonstrate the comparison of $H_{0}$ measurement for different combination of data sets with results of Planck 2018 and R22 for kaniadakis entropy  without neutrinos

	\begin{table*}[h]
		\centering
		\caption{Comparison  $H_0$, $H_0$ Tension for different combinations of data for Kaniadakis with neutrinos}
		\begin{tabular}{|l|c|c|c|c|}
			\hline
		dataset	& $H_{0}$ ($\mathrm{km \, s^{-1} \, Mpc^{-1}}$) & Tension with R22 & Tension with Planck 2018 \\
			\hline
			CMB  & $69.3\pm 1.7$ &  $2.13\sigma$ & $0.033\sigma$ \\
			CMB + Pantheon & $71.68\pm 2.2$ &  $0.89\sigma$ & $1.89\sigma$ \\
			CMB + CC & $70.97\pm 1.6$ &  $1.43\sigma$ & $2.12\sigma$ \\
			CMB + BAO & $71.56\pm 1.51$ &  $1.91\sigma$ & $2.61\sigma$ \\
			CMB + BAO + Pantheon + CC & $70.61\pm 1.49$ &  $1.66\sigma$ & $2.04\sigma$ \\
			\hline
		\end{tabular}
	\end{table*}
	
	\begin{table*}[h]
		\centering
		\caption{Comparison  $S_8$, and  $S_8$ Tension for different combinations of data for Kaniadakis with neutrinos}
		\begin{tabular}{|l|c|c|c|c|c|c|c|}
			\hline
		dataset	& $S_{8}$ & Tension with kiDS & Tension with Planck & Tension with DES \\
			\hline
			CMB & $0.801 \pm 0.067$ & $0.51\sigma$ & $0.45\sigma$ & $0.36\sigma$ \\
			CMB + Pantheon& $0.788 \pm 0.047$ & $0.45\sigma$ & $0.9\sigma$ & $0.24\sigma$ \\
			CMB + CC & $0.788 \pm 0.031$ & $0.15\sigma$ & $1.3\sigma$ & $0.33\sigma$ \\
			CMB + BAO & $0.789 \pm 0.049$ & $0.45\sigma$ & $0.84\sigma$ & $0.25\sigma$ \\
			CMB + BAO + Pantheon + CC & $0.788 \pm 0.032$ & $0.65\sigma$ & $1.27\sigma$ & $0.33\sigma$ \\
			\hline
		\end{tabular}
	\end{table*}

\begin{table}[h]
		\centering
\caption{Cosmological Parameter Results for Different Datasets for Kaniadakis horizon entropy with neutrinos}
\resizebox{0.9\textwidth}{!}{	
	\begin{tabular}{|l|c|c|c|c|c|}
		\hline
		{Parameter} & {CMB+Pantheon} & {CMB+CC} & {CMB+BAO} & {CMB+ALL} & {CMB}\\
		\hline
		${\Omega_b h^2}$ & $0.02226 \pm 0.00023$ & $0.02224 \pm 0.00024$ & $0.02238 \pm 0.00019$ & $0.02221 \pm 0.00022$ &$0.02236 \pm 0.0002$ \\
			\hline
		${\Omega_c h^2}$ & $0.1202 \pm 0.0037$ & $0.1200 \pm 0.0037$ & $0.1185 \pm 0.0030$ & $0.1190 \pm 0.0028$ &$0.1185 \pm 0.0032$\\
			\hline
		${100\theta_{MC}}$ & $1.04088 \pm 0.00056$ & $1.04089 \pm 0.00057$ & $1.04111 \pm 0.00044$ & $1.04099 \pm 0.00050$ &$1.04112 \pm 0.00043$\\
			\hline
		${\tau}$ & $0.0547 \pm 0.0076$ & $0.0544 \pm 0.0076$ & $0.0552 \pm 0.0080$ & $0.0552^{+0.0054}_{-0.0078}$ &$0.05526 \pm 0.0081$\\
			\hline
		${{\rm{ln}}(10^{10} A_s)}$ & $3.045 \pm 0.018$ & $3.044 \pm 0.018$ & $3.042 \pm 0.019$ & $3.043^{+0.016}_{-0.019}$&$3.042 \pm 0.019$ \\
			\hline
		${n_s}$ & $0.9686 \pm 0.0085$ & $0.9677 \pm 0.0087$ & $0.9649 \pm 0.0073$ & $0.9657 \pm 0.0071$&$0.9645 \pm 0.0078$  \\
	    	\hline
			${\Sigma_{m_{\nu}}}(\rm eV)$ & $0.146$ & $0.126$ & $0.116 $ & $0.113 $ &$0.276$\\
		\hline
	\end{tabular}
}
\end{table}

\begin{figure}
	\includegraphics[width=8.5 cm]{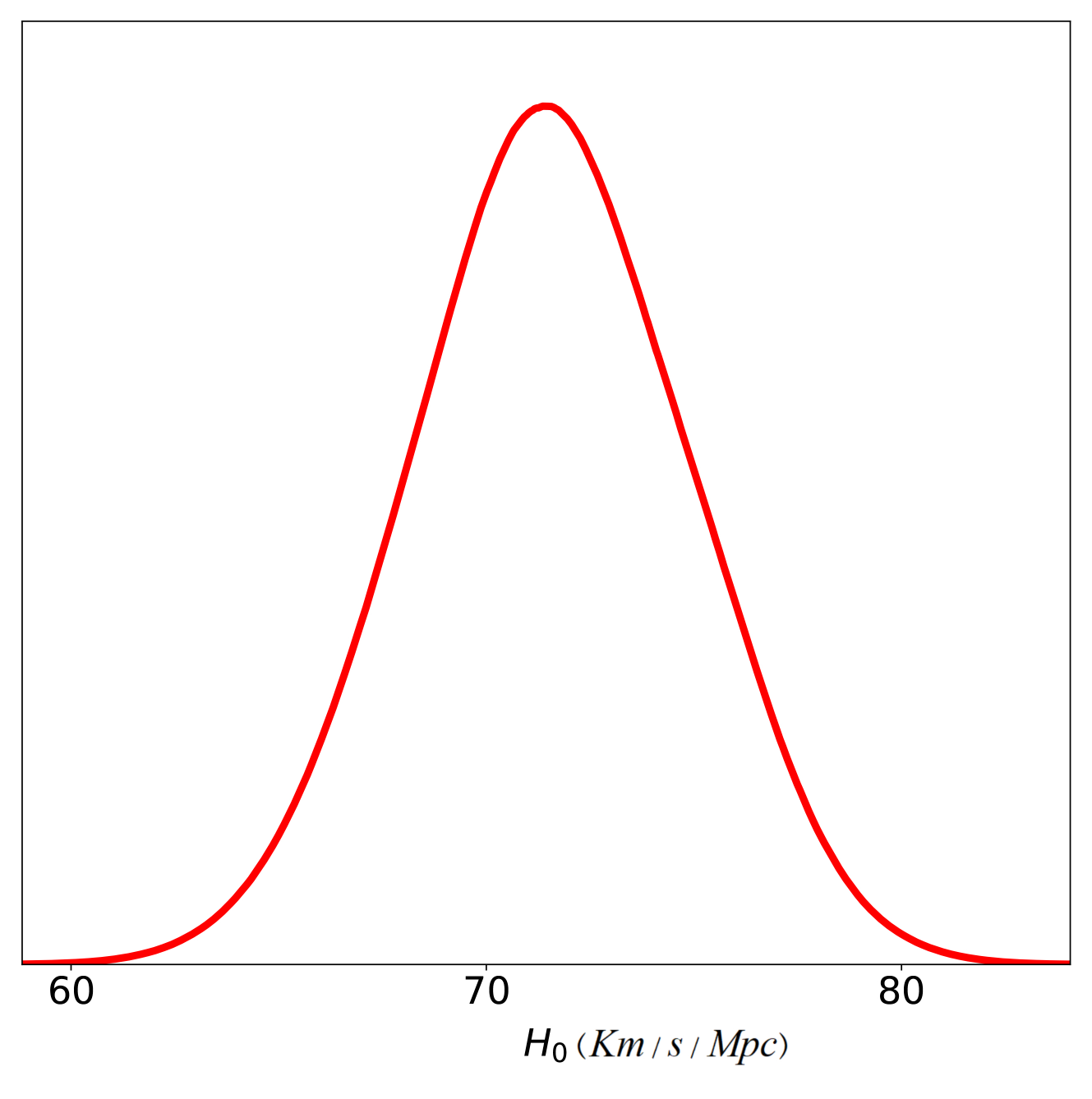}
	\vspace{-0.12cm}
	\caption{\small{Comparison  results of the  $H_{0}$ for  combination dataset(CMB + BAO + Pantheon + CC) for kaniadakis entropy  with neutrinos. }}\label{fig:omegam2}
\end{figure}
\begin{figure}
	\includegraphics[width=8.5 cm]{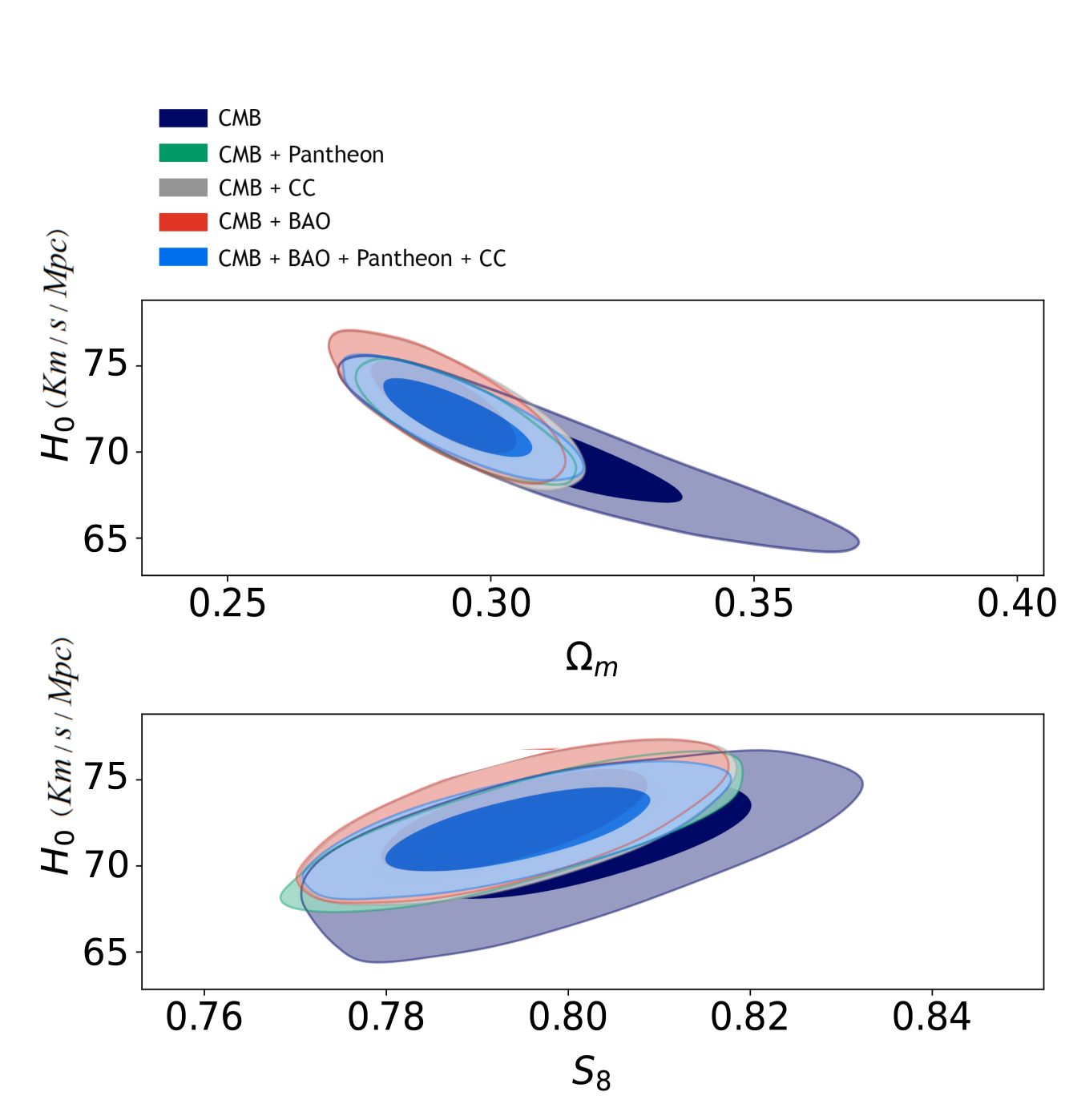}
	\vspace{-0.12cm}
	\caption{\small{Comparison  results of the  $\Omega_{m}$ , $S_{8}$ according to $H_{0}$ for different combination dataset for kaniadakis entropy  with neutrinos. }}\label{fig:omegam2}
\end{figure}
\begin{figure*}
	\includegraphics[width=16.5 cm,height=18 cm]{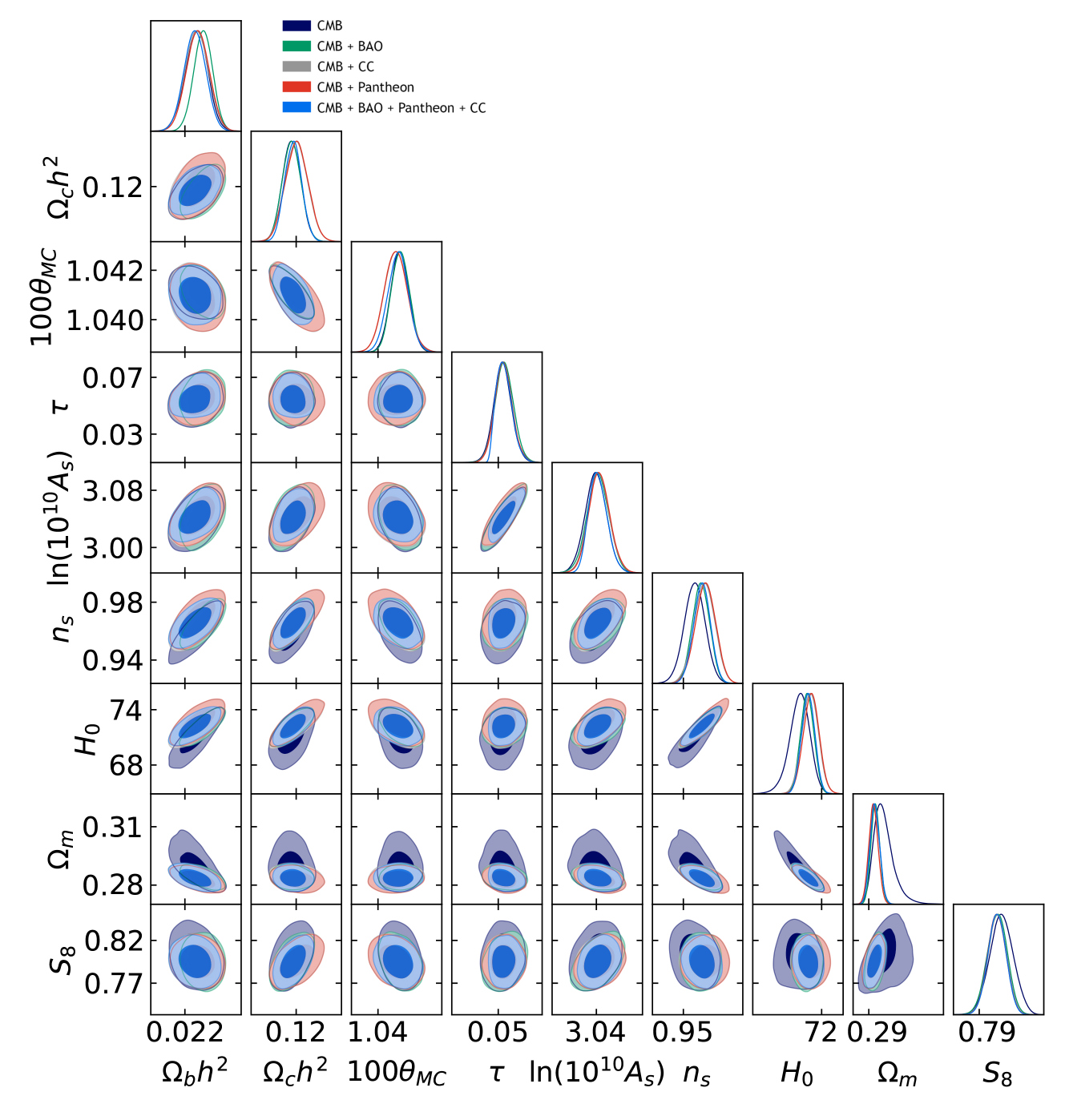}
	\vspace{-0.12cm}
	\caption{\small{Comparison  results of the  $S_{8} $ , $\Omega_b h^2   $, $\Omega_c h^2   $, $\tau$, $n_s  $, ${\rm{ln}}(10^{10} A_s)$, $100\theta_{MC} $, $H_{0}(km/s/Mpc
			)$ for different combination dataset for kaniadakis entropy  with neutrinos. }}\label{fig:omegam2}
\end{figure*}
\begin{figure*}
	\includegraphics[width=16.5 cm,height=10 cm]{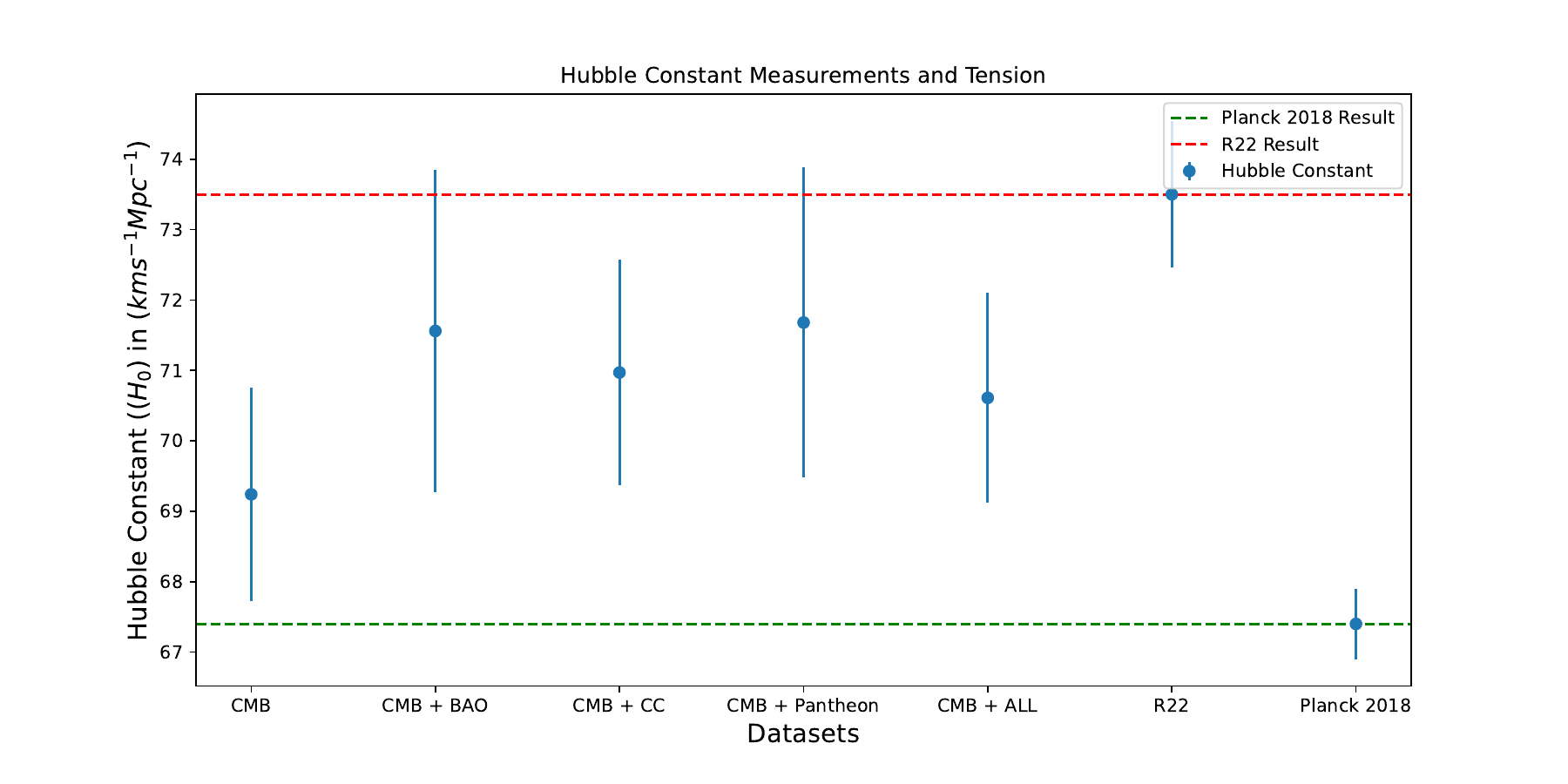}
	\vspace{-0.12cm}
	\caption{\small{The comparison of $H_{0}$ measurement for different combination of data sets with results of Planck 2018 and R22 for kaniadakis entropy  with neutrinos. }}\label{fig:omegam2}
\end{figure*}

As we have observed, when we use Kaniadakis entropy to consider the Hubble tension and $ \sigma_{8} $ tension without taking neutrinos into account, the results obtained are very similar to Planck's 2018 findings. However, when we include the neutrino term, the Hubble and $ \sigma_{8} $ tensions are alleviated, and the  value of $ \sigma_{8} $ is in full agreement with the value obtained in kiDS and DES.

The modified Friedmann equations resulting from the Kaniadakis horizon entropy with neutrinos include an extra term that can be seen as an effective dark energy term. This term is directly proportional to the Kaniadakis parameter K, which is similar to the one found in the Kaniadakis horizon entropy without neutrinos. However, in the Kaniadakis horizon entropy with neutrinos, the value of K may be influenced by the neutrino density. Including neutrinos in the model can impact the value of K because they can interact with other particles, leading to energy transfer between different parts of the universe. This interaction can cause the neutrino distribution to become anisotropic, which can ultimately affect the value of K.

 Furthermore, in the following, we put a constraint on k  in each scenario (kaniadakis entropy without neutrinos and with neutrinos), and we obtain $ k = 0.12\pm 0.41 $ for kaniadakis entropy without neutrinos and $ k = 0.39\pm 0.4 $ for kaniadakis entropy in the presence the neutrinos. The k results for both scenarios are plotted in Fig. 11.
 Figure 11 presents a comprehensive comparison of the parameter k in two distinct scenarios: one utilizing Kaniadakis entropy without neutrinos and the other incorporating neutrinos into the entropy model. The parameter 
 k holds significance in characterizing the shape and behavior of the entropy distribution. The juxtaposition of these scenarios allows for a detailed examination of how the inclusion of neutrinos influences the value of 
 k and, consequently, the overall entropy dynamics. This comparison sheds light on the nuanced interplay between neutrinos and entropy within the Kaniadakis framework, providing valuable insights into the role of neutrinos in shaping the entropy landscape.

\begin{figure}
	\includegraphics[width=8.5 cm]{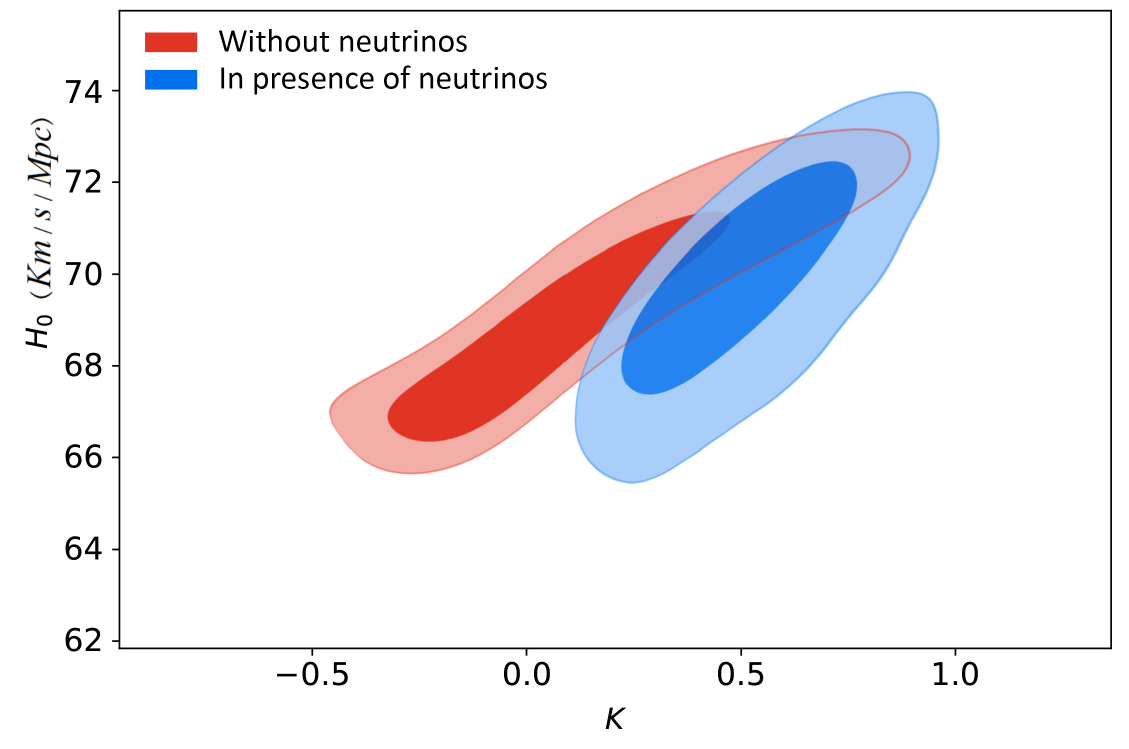}
	\vspace{-0.12cm}
	\caption{\small{Comparison  results of the  k parameter for both scenario (kaniadakis entropy Without neutrinos and with neutrinos) }}\label{fig:omegam2}
\end{figure}

\begin{table*}
	\tiny
	\caption{{\small  Mean values of free parameters of various models with 1$ \sigma $ error bar for combinations data}}
	\begin{center}
		\resizebox{0.8\textwidth}{!}{  
			\begin{tabular}{ c |c c c c c c c c c  } 
				\hline
				\hline
				Models & $\chi^2_{\rm  CMB + Pantheon}$ & $\chi^2_{\rm  CMB + CC}$  & $\chi^2_{\rm  CMB+BAO }$&$\chi^2_{\rm  ALL}$ \\ 
				\hline
				\hline
				$\Lambda$CDM  & $3585.821$ & $2791.634$ & $2776.083$ & $3614.172$   \\
				First scenario& $3581.874$ & $2774.114$ & $2768.917$ & $3605.766$   \\
				Second scenario & $3580.824$ & $2774.038$ & $2766.284$ & $3602.091$   \\
				\hline
				\hline
			\end{tabular}
		}
	\end{center}
	\label{table_chi}
\end{table*}

\begin{table*}
	\caption{{\small  Mean values of free parameters of various models with 1$ \sigma $ error bar for combinations data}}
	\begin{center}
		\resizebox{1\textwidth}{!}{  
			\begin{tabular}{ c |c c c c c c c c c  } 
				\hline
				\hline
				Models & $ \Omega_{DE}$ & $ \Omega_{\rm m}$  & $ \Omega_{\Lambda}$&$k$&$ \Omega_{\nu}$&$ H_{0}(km/s/Mpc
				) $& AIC \\ 
				\hline
				\hline
				$\Lambda$CDM  & $-$ & $0.312\pm0.2$ & $0.678\pm0.2$ & $-$ & $-$ & $67.8\pm1.1$ & $3620.172$   \\
				First scenario& $0.69\pm 0.21$ & $0.303\pm 0.0077$ & $-$ & $0.12\pm0.41$ & $-$ & $69.07\pm 1.51$ & $ 3613.766 $  \\
				Second scenario & $0.706\pm0.29$ & $0.281 ^{+0.036}_{-0.059}$ & $-$ & $0.39\pm 0.4$ &$0.0027\pm0.0009$ & $70.61\pm1.49$ & $3612.091$   \\
				\hline
				\hline
			\end{tabular}
		}
	\end{center}
	\label{table_chi}
\end{table*}

\section{Conclusion}

In this study, we explored two distinct scenarios aimed at addressing the Hubble tension and \( \sigma_8 \) tension. In the first scenario, we determined the values of \( h \) and \( \Omega_m \) utilizing the Friedman equations within the framework of Kaniadakis entropy. Subsequently, we derived \( H_0 \) and \( \sigma_8 \) based on these parameter estimates. In the second scenario, we introduced the neutrino term into the Friedman equations and repeated the parameter calculations.

For the combination dataset (CMB+BAO+CC+Pantheon) in the first scenario, we obtained \( H_0 = 69.07 \pm 1.51 \, \text{km s}^{-1} \text{Mpc}^{-1} \) at 68\% CL, closely aligning with Planck 2018 results. Notably, this result exhibits a 1.04\( \sigma \) tension with Planck, a 2.39\( \sigma \) tension with R22. The derived \( S_8 \) value is \( 0.802\pm 0.043 \). When comparing this outcome with DES, kiDS, and Planck, the tensions are 0.56\( \sigma \), 0.81\( \sigma \), and 0.66\( \sigma \), respectively.

In the second scenario, we found \( H_0 = 70.61 \pm 1.49 \, \text{km s}^{-1} \text{Mpc}^{-1} \) at 68\% CL, once again in proximity to Planck 2018 results. However, this result exhibits a 2.04\( \sigma \) tension with Planck, and a 1.66\( \sigma \) tension with R22. The corresponding \( S_8 \) value is \( 0.788\pm 0.032 \). Notably, this outcome is in total agreement with kiDS, with a 0.65\( \sigma \) . Additionally, there is a 1.27\( \sigma \) tension with Planck and a 0.33\( \sigma \) tension with DES.

As observed, when considering the Hubble tension and \( \sigma_{8} \) tension using Kaniadakis entropy without the presence of neutrinos, the results closely approximate Planck's 2018 results. Intriguingly, upon incorporating the neutrino term, the Hubble and \( \sigma_{8} \) tensions are alleviated, and the \( \sigma_{8} \) value aligns entirely with the value obtained in kiDS.

Furthermore, we imposed a constraint on \( k \) in each scenario, both for Kaniadakis entropy without neutrinos and with neutrinos, yielding \( k = 0.12 \pm 0.41 \) for Kaniadakis entropy without neutrinos and \( k = 0.39 \pm 0.4 \) for Kaniadakis entropy in the presence of neutrinos.
We found that the best fit was for $\Lambda$ = 0.0052. This value suggests that $\Lambda$ has little effect as dark energy, and that the k parameter can serve as a substitute for dark energy. Therefore, we can conclude that Kaniadakis horizon entropy is capable of explaining the accelerated expansion of the universe without the need for dark energy.
Finally, as illustrated in Tables 9,10 the second scenario demonstrates a better fit than the first scenario, and both outperform \( \Lambda \)CDM for the complete dataset combination, leading to an improvement in the \( \chi^{2} \) statistic. Consequently, the inclusion of the neutrino term emerges as particularly impactful, effectively mitigating tensions. This is underscored by the notable improvement in \( \chi^{2} \), indicating that the addition of the neutrino term provides a robust constraint on the density of matter. In conclusion, our findings suggest that the inclusion of the neutrino term plays a pivotal role in alleviating tensions and enhancing the overall fit of the model to the observational data.

\end{document}